\documentclass[prereprint]{revtex4}
\usepackage{graphicx}
\usepackage{enumitem}

\usepackage{mathtools}
\usepackage{graphicx}   
\usepackage{color}
\usepackage{ulem}
\usepackage{caption}
\usepackage{comment}

\usepackage{array}
\newcolumntype{C}[1]{>{\centering\arraybackslash}p{#1}} 
\usepackage{booktabs}
\usepackage{multirow}

\usepackage{booktabs}

\begin{document}
\title{Reaction coordinates and rate constants for liquid droplet nucleation: quantifying the interplay between driving force and memory}

	\author{Sun-Ting Tsai}
 \affiliation{Department of Physics and Institute for Physical Science and Technology,
 University of Maryland, College Park 20742, USA.}

	\author{Zachary Smith}
 \affiliation{Biophysics Program and Institute for Physical Science and Technology,
 University of Maryland, College Park 20742, USA.}

 \author{Pratyush Tiwary*}
 \affiliation{Department of Chemistry and Biochemistry and Institute for Physical Science and Technology,
 University of Maryland, College Park 20742, USA.}
 
	\date{\today}
	
  \begin{abstract}
In this work we revisit the classic problem of homogeneous nucleation of a liquid droplet in a supersaturated vapor phase. We consider this at different extents of the driving force, which here is the extent of supersaturation, and calculate a reaction coordinate (RC) for nucleation as the driving force is varied. The RC is constructed as a linear combination of three order parameters, where one accounts for the number of liquid-like atoms, and the other two for local density fluctuations. The RC is calculated from all-atom biased and unbiased molecular dynamics (MD) simulations using the spectral gap optimization approach ``SGOOP" [P. Tiwary and B. J. Berne, Proc. Natl. Acad. Sci. U. S. A. 113, 2839 (2016)]. Our key finding is that as the supersaturation decreases, the RC ceases to simply be the number of liquid-like atoms, and instead it becomes important to explicitly consider local density fluctuations that correlate with shape and density variations in the nucleus. All three order parameters are found to have similar barriers in their respective potentials of mean force, however, as the supersaturation decreases the density fluctuations decorrelate slower and thus carry longer memory. Thus at lower supersaturations density fluctuations are non-Markovian and can not be simply ignored from the RC by virtue of being noise. Finally, we use this optimized RC to calculate nucleation rates in the infrequent metadynamics framework, and show it leads to more accurate estimate of the nucleation rate with four orders of magnitude acceleration relative to unbiased MD.
\end{abstract}


	\maketitle
\section{Introduction}
The nucleation of one phase from another is considered as the first step of several phase transitions in chemical physics, with relevance to diverse and important problems in science and technology.\cite{bartels2013chemistry,murray2010heterogeneous,mazur1970cryobiology,cox2007selective,erdemir2007polymorph,sloan2003fundamental,hammerschmidt1934formation,harper1997atomic,cohen2013proliferation} Through experiments, simulations and theory, this problem has been extensively studied over the decades.\cite{sosso2016crystal} In spite of so much attention being lavished upon this problem, it continues to be a difficult challenge. For instance, the typical size of the critical nucleus is so small that it becomes difficult to observe and analyze it through experiments, or model through continuum based theories. The tens to hundreds of atoms size of the nucleus thus makes it in principle ideal for probing through all-atom molecular dynamics (MD) simulations. However, this is easier said than done due to the inherent rare event nature of the problem, where one nucleation event can take seconds, hours or longer, making it far beyond the microsecond timescale available through the fastest supercomputers. This has led to the development of a plethora of  sampling schemes that attempt to enhance the process of nucleation in a controllable manner.\cite{torrie1977nonphysical,kumar1992weighted,meta_laio,wtm,van2003novel,moroni2004investigating,allen2006simulating,allen2009forward} These various sampling methods need the pre-determination of slow degree or degrees of freedom relevant to the nucleation process being studied. This slow degree of freedom which is most informative of the underlying physics is referred to as the reaction coordinate (RC).\cite{berezhkovskii2005one,sgoop,szabo_anisotropic} In sampling methods such as metadynamics,\cite{wtm, arpc_meta} where one gradually deposits a time-dependent bias to escape free energy minimum, the need to know a reasonably good RC beforehand is well-documented. In a different class of methods such as forward flux sampling (FFS),\cite{sarupria2012homogeneous} recent work has started to highlight how FFS can benefit from pre-knowledge of adequate slow order parameters or the RC.\cite{defever2019contour} Finally in methods such as transition path sampling (TPS)\cite{dellago1998efficient,bolhuis1998sampling} and variants thereof,\cite{van2003novel,moroni2004investigating} this dependence on pre-knowledge of RC is somewhat mitigated, but instead one becomes reliant on the accuracy of the initial path used in the sampling. In any case, one can say with confidence that any sampling scheme for the study of nucleation can only benefit from a prior sense of an approximate RC for nucleation, with the degree of benefit varying from scheme to scheme.

In this work, we consider what is arguably the simplest of nucleation problems, namely that of the homogeneous nucleation of a liquid droplet in a supersaturated vapor phase at different supersaturation levels\cite{chkonia2009evaluating}. The system is modeled using Lennard-Jones interactions.\cite{salvalaglio_argon, chkonia2009evaluating} Even in this simplest of problems, we find that the RC for homogeneous nucleation deviates significantly from standard assumptions made so far in theoretical and simulation approaches.\cite{salvalaglio_argon,kuipers2009non,ford2004statistical} Our calculations of the RC are performed using a spectral gap based optimization method ``SGOOP", originally proposed by Tiwary and Berne \cite{sgoop,anisod_sgoop,sgoop_fullerene,multisgoop}, for the automatic construction of RC from different trial order parameters.  We find that there exists a supsersaturation (or more generally speaking, the driving force for nucleation) dependent interplay between size, density and shape of the nucleus. This interplay leads to a non-trivial RC that goes far beyond a spherical, uniformly dense nucleus assumed in classical nucleation theory (CNT).\cite{weber1926keimbildung,farkas1927keimbildungsgeschwindigkeit,becker1935kinetische,zeldovich1943theory,maris2006introduction,Kalikmanov2013} While we define RC more rigorously in the main text, here we summarize it as a low-dimensional variable permitting a Markovian description of the underlying high-dimensional dynamics.\cite{szabo_timescale,berezhkovskii2005one,kuipers2009non} Our key finding is that as the supersaturation decreases, the RC becomes composed of not just the number of atoms in the largest liquid-like cluster, but it also becomes helpful to consider the spatial fluctuations of the aforementioned quantity. These fluctuations display similar barriers as the number of liquid-like atoms, but have a longer autocorrelation time (or equivalently, slower diffusion). This diffusion anisotropy becomes stronger as the supersaturation decreases. In the spirit of works by Szabo, Peters, Hynes and others,\cite{van1982reactive,berezhkovskii2005one,hynes1985chemical,peters2013reaction} we find that the RC itself starts to align with the direction of slowest diffusion or longest memory, given that the free energy barriers in the directions of various individual order parameters are similar. Finally, we use the optimized RC as a biasing variable in infrequent metadynamics calculations,\cite{meta_time} which allow recovering unbiased kinetic rate constants from biased simulations. We find that considering this diffusion anisotropy adjusted RC in infrequent metadynamics leads to more accurate estimates of the nucleation rate across different supersaturations with orders of magnitude speed-up relative to unbiased MD. 

We believe that our work demonstrates the potential of using methods such as SGOOP in unraveling the subtle aspects of the RC in complex nucleation problems. Such a RC first of all directly gives useful physical insight into the processes at play, but secondly, as we show here it also serves as useful descriptor for performing enhanced sampling simulations including metadynamics and beyond.

\section{Theory and Methods}

\subsection{Order parameters}
In order to motivate this work and the various order parameters we consider here, we start with a brief description of CNT which has been a basic building block in the study of nucleation. In CNT, the first liquid droplet formed in the vapor is treated as spherical shaped and uniformly dense.\cite{weber1926keimbildung,farkas1927keimbildungsgeschwindigkeit,becker1935kinetische,zeldovich1943theory,maris2006introduction,Kalikmanov2013} The nucleation process is then modeled by balancing the surface tension penalty with chemical potential benefit.\cite{rao1978computer} This simple theory, though it captures qualitatively how nucleation happens, however fails to quantify the true nucleation rate in any practical sense. It is believed that CNT makes several oversimplified assumptions especially incorrectly assuming that the cluster is spherical and uniform.\cite{wang2007homogeneous,kalikmanov2008argon} By using numerical and experimental tools, the lack of sphericity and uniform density has indeed been documented in crystal nucleation and in nucleation in more complex systems.\cite{ten1997enhancement,trudu2006freezing,moroni2005interplay,kwon2015heterogeneous,zhou2019observing} However such simulations and experiments are expensive, and it has been hard to quantitatively probe such effects even in the simple gas system such as the one used in this work. 

A popular order parameter that goes beyond the spherical nucleus approximation of CNT was introduced by Frenkel and ten Wolde.\cite{ten1998computer} This order parameter $n$ equals the number of liquid-like atoms in the system in a way that it is still a continuous and differentiable function\cite{ten1998computer} of atomic coordinates, a necessity for the biased simulations we perform later. In this definition, an atom is classified as liquid if it has more than 5 neighboring atoms. The number of the neighborhood atoms of the atom with label $i$, or equivalently the coordination number $c_{i}$, is calculated through the use of a switching function as follows:
\begin{align}
c_{i}=\sum_{j\neq i}\frac{1-(r_{ij}/r_{c})^{6} }{1-(r_{ij}/r_{c})^{12} }
\end{align}
where the summation is carried over all atoms $j\neq i$. The distance $r_{ij}$ between atoms $i$ and $j$ needs to be less than a cut-off $r_{c}$ to be considered as neighbors. The number of liquid phase atoms $n$ is then calculated using a similar form with threshold value $c_{l}$ which we take to be 5 here in spirit of Ref. \onlinecite{ten1998computer}:
\begin{align}
n=\sum_{i=1}^{N}\frac{1-(c_{l}/c_{i})^{6} }{1-(c_{l}/c_{i})^{12} }
\end{align}

\begin{figure}[!h]
  \centering
  \includegraphics[width=8cm]{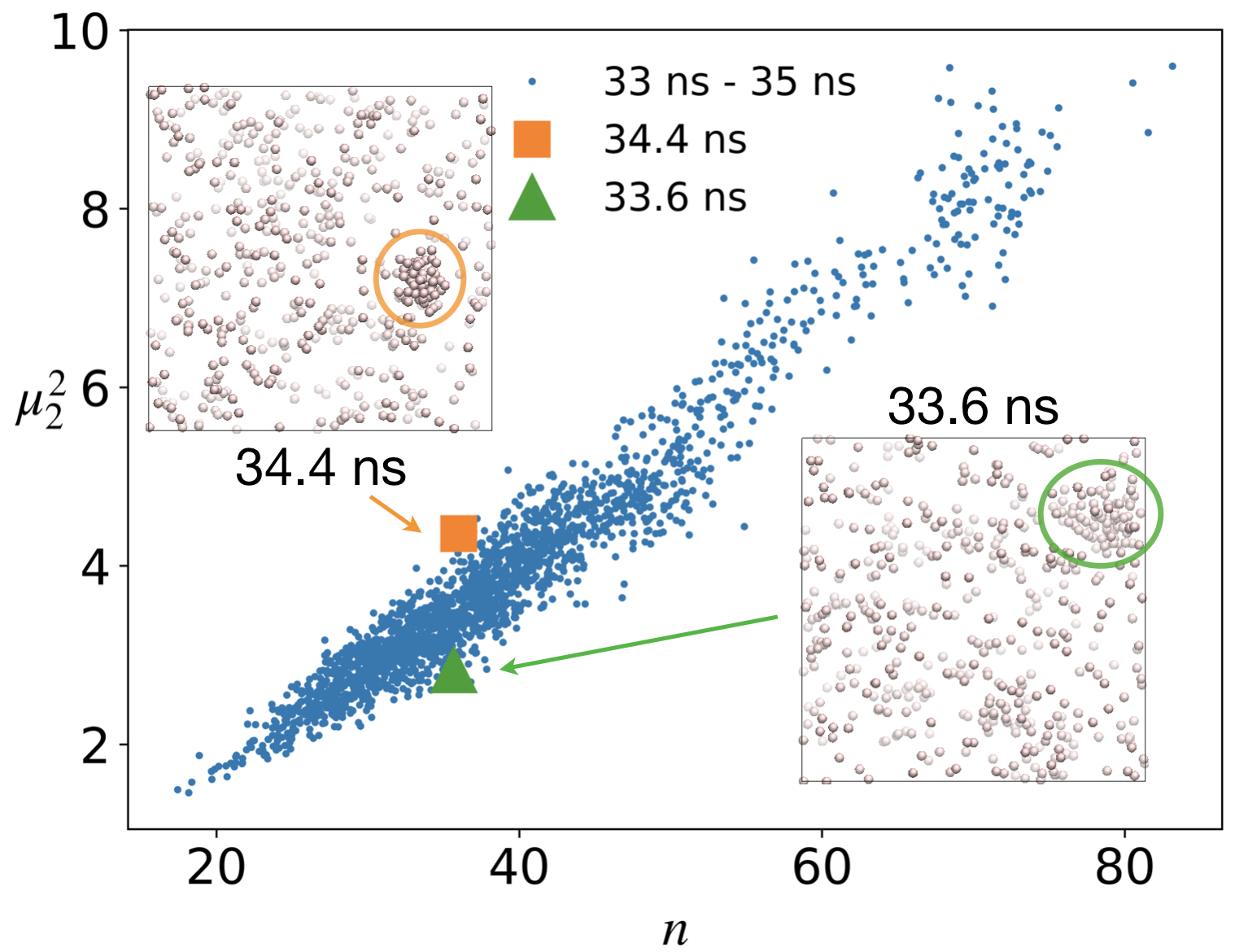}
  \caption
  { Here we show an unbiased MD trajectory in ($n, \mu^{2}_{2}$) space at supersaturation 11.43, for a 2 $ns$ interval between 33 $ns$ to 35 $ns$. The panels are the snapshots at similar $n$ but different $\mu^{2}_{2}$, showing clearly how there can be clusters with same $n$ but otherwise very different properties including density an compactness. For instance here, at higher $\mu^{2}_{2}$, the cluster is visibly more compact than the one at lower $\mu^{2}_{2}$.
}\label{nuclei}
\end{figure}

The above defined $n$ thus captures the total number of liquid-like atoms in the system\cite{ten1998computer}. It is however oblivious to details such as the density of the clusters in which these atoms are present, if there are more than 1 clusters, the shape of these clusters and other nuances. In order to consider these, we propose including the second and third moments of the distribution of coordination numbers, defined as $\mu^{2}_{2}$ and $\mu^{3}_{3}$ respectively. \cite{moroni2005interplay,tribello2017analyzing,tribello2010self} These are explicitly calculated as:
\begin{align}
\mu^{2}_{2}=\frac{1}{N}\sum_{i=1}^{N}(c_{i}-\overline{c})^{2},\quad\mu^{3}_{3}=\frac{1}{N}\sum_{i=1}^{N}(c_{i}-\overline{c})^{3}
\end{align}
where $\overline{c}$ is the average of the coordination number. In Fig. \ref{nuclei} we show a representative unbiased MD trajectory in $(n,\mu_2^2)$ space where it can be seen that at roughly the critical size of $n=30$ different $\mu^{2}_{2}$ can represent clusters with strikingly different profiles in terms of shape, density and compactness.

The distribution of the coordination numbers captures the variations in density of the liquid phase and thus can be used to study the local properties of the liquid droplets. Another nice feature of these moments is that they can easily be generalized for multi-component systems to take into account the variation in density related to specific species.\cite{ten1997enhancement,kwon2015heterogeneous,kathmann2004multicomponent} Our RC is then expressed as a linear combination of these three order parameters $n$, $\mu^{2}_{2}$ and $\mu^{3}_{3}$. 

\subsection{Nucleation rate}
The process of nucleation is inherently stochastic in nature and satisfies the law of rare events. In other words, different independent observations of nucleation should give a distribution of nucleation times adhering to a Poisson process.\cite{resnick2013adventures} If we let $P(t)$ denote the survival probability of not observing any liquid droplets at and until time $t$, it will satisfy the following relation valid for all Poisson processes:
\begin{align}
P(t)=e^{-t/\tau}
\label{poisson_eq}
\end{align}

where $\tau$ is the characteristic time for the first nucleation event, the inverse of which can be interpreted as the nucleation rate. In this work, we find the characteristic time by performing multiple independent simulations starting from the system in gaseous state with randomized velocities (other simulation details in Sec. \ref{setup}), and collect the statistics of transition times until the first nucleation event. The nucleation time is then obtained by performing a Poisson fit to these independent observations following the protocol outlined in Ref. \onlinecite{salvalaglio_argon,pvalue}.

\subsection{Simulation set-up}
\label{setup}
The simulations were performed under the constant number, volume, temperature (NVT) ensemble with N=512 argon atoms and average temperature fixed at 80.7 K. The volume of the simulation box was set in order to correspond to desired supersaturation levels $S$ detailed in Table \ref{WTmetaD_para}. In order to compare our results with unbiased nucleation rates in Ref. \onlinecite{chkonia2009evaluating}, our cubic box size ranged from 9.5 nm to 11.5 nm. The interaction between atoms were modeled through a Lennard-Jones potential with $\epsilon=0.99797$ kJ/mol and $\sigma=0.3405$ nm \cite{chkonia2009evaluating}. The potential was truncated with cutoff at 6.75 $\sigma$. The velocity rescale thermostat with time constant of 0.1 $ps$ was used to do temperature coupling. \cite{bussi2007canonical}  All simulations were performed using GROMACS version 2016.5\cite{lindahl2001gromacs} patched with PLUMED version 2.4.2.\cite{plumed2019nature}

\subsection{Reaction coordinate}
\label{sec_rc}
In order to quantify how these various order parameters $n$, $\mu^{2}_{2}$, and $\mu^{3}_{3}$ matter for the process of nucleation, we intend to learn a RC $\chi$ as their linear combination. In addition to quantifying exactly how much these order parameters matter for driving nucleation, this RC will also serve as a crucial input for biased simulations to be performed later in this work. We first carefully define what exactly we mean by RC. 

The RC for a given molecular system is traditionally defined as an abstract low-dimensional coordinate that best captures progress along relevant reaction pathway. While this intuitive notion can be formalized and quantified in several different ways, here we use the definition of RC as follows. For a given multidimensional complex system undergoing a certain dynamics, it is an optimal low-dimensional variable such that the multidimensional dynamics of the full system in terms of movement between different metastable states can be mapped into Markovian dynamics between various states viewed as a function of the RC\cite{berezhkovskii2005one,berezhkovskii2011time}. Thus an optimal RC is a low-dimensional mapping which best satisfies (i) thermodynamic truthfulness: demarcating between the various relevant metastable states present in the actual high-dimensional system, (ii) kinetic truthfulness: preserving pathways for moving between these different states, and (iii) timescale separation: displaying a clean-cut separation of timescales between the relaxation times in the various metastable states, and the time spent in the actual event of crossing from one state to another. 

\subsection{SGOOP}
\label{sec_sgoop}
 To find such a RC, here we use the method ``Spectral gap optimization of order parameters (SGOOP)". This method uses the principle of maximum caliber (``MaxCal"), which is similar to path entropy,\cite{dixit2015inferring,jaynes_caliber,caliber1,dixit2018perspective} to construct a transition probability matrix along any candidate RC, and then calculates its eigenvalues $\lambda_{0}=1>\lambda_{1}>\lambda_{2}> ... $. Here $\lambda_{0}=1$ corresponds to stationary state, while the other eigenvalues carry information about the timescales of various dynamical processes. The best RC will then produce a transition matrix $K$ with a maximal timescale separation between visible slow and hidden fast processes. This timescale separation, also known as spectral gap, is quantified as the difference $\lambda_{n} - \lambda_{n+1}$, where $n$ is the number of discernible energy wells along the putative RC. SGOOP needs two key inputs: (i) an estimate of the stationary probability density $\pi$ along any putative RC, and (ii) some dynamical observables or constraints. With these inputs, the SGOOP transition matrix $K$ can be formulated as follows:
\begin{align}
K_{mn}=\Lambda\sqrt{\frac{\pi_{n}}{\pi_{m}}}
\label{eq_maxcal1}
\end{align}
where $\pi_{m}$ is the stationary probability along any putative, spatially-discretized RC $\chi$ with $m$ denoting the grid index, and $\Lambda$ is a dynamical observable we will revisit shortly. $K_{mn}$ gives the rate for moving from grid $m$ to grid $n$ in a small time interval. The input (i), namely the stationary density $\pi$ can come from unbiased MD at high enough supersaturations, or if the supersaturation is too low to permit unbiased MD, it can come through the use of preliminary metadynamics along a trial RC, followed by reweighting.\cite{tiwary_rewt} For input (ii), namely calculation of the dynamic observable needed to constrain the maximum caliber estimate of rate matrix, we run short unbiased MD runs which calculate the mean number of nearest neighbor transitions $\langle N\rangle$ along any putative RC. It is then easy to show\cite{tiwary2017predicting} that the dynamical observable $\Lambda$ in Eq. \ref{eq_maxcal1} is given by
\begin{align}
\Lambda=\frac{\langle N\rangle}{\sqrt{\pi_{m}\pi_{n}}}
\label{maxcal2}
\end{align}
Equivalently\cite{tiwary2017predicting}, if one was to completely by-pass the MaxCal framework, a very similar equation as Eq. \ref{eq_maxcal1} can be derived by comparing a master equation along $\chi$ with a discretized Smoluchowski equation along the same. \cite{szabo_bicout} Then the prefactor $\Lambda$ becomes:
\begin{align}
\Lambda=\frac{D_{\chi}}{2d^{2}}
\label{rate_diffusion}
\end{align}
where $D_{\chi}$ is the position-dependent diffusivity along the coordinate $\chi$ and $d$ is the grid spacing along $\chi$. 

Eqs. \ref{eq_maxcal1}--\ref{rate_diffusion} collectively show that the rate matrix $K$, hence the spectral gap, and consequently the optimized RC, depend not just on the free energy barriers that would be encapsulated in the stationary density $\pi$, or equivalently in the associated free energy, but that the dynamics of the system as captured in the diffusivity of the various order parameters can also play a significant role in the RC. As we will show later in Sec. \ref{results}, we find this to be a very important point in the context of liquid droplet nucleation.

\subsection{Metadynamics}

\begin{table}
\renewcommand{\arraystretch}{1.25}
\begin{tabular}{|c|c|c|c|c|c|c|c|}
\hline
\multicolumn{8}{|c|}{WTmetaD parameters} \\
\hline
Label & $S$ & $L$ (nm) & $h$ (kJ/mol) & $\omega_{n}$ & $\omega_{\chi}$ & $\Delta t$ ($ps$) & $\gamma$ \\ \hline
$S_{1}$ & 13.65 & 9.9 & 0.01 & 0.5 & 0.08 & 25 & 5 \\ \hline
$S_{2}$ & 12.80 & 10.1 & 0.05 & 0.5 & 0.08 & 25 & 8 \\ \hline
$S_{3}$ & 11.43 & 10.5 & 0.2 & 0.5 & 0.08 & 25 & 8 \\ \hline
$S_{4}$ & 9.87 & 11.0 & 0.2 & 0.5 & 0.08 & 25 & 8 \\ \hline
$S_{5}$ & 9.04 & 11.3 & 0.2 & 0.5 & 0.08 & 25 & 8 \\ \hline
\end{tabular}
\caption{The metadynamics parameters used at different supersaturation levels: $S$ represents the supersaturation level, $L$ is the size of simulation cubic box which we used to control the supersaturation. Gaussian bias kernels of starting height $h$ and width $\omega$ were added every $\Delta t$, which was kept same for metadynamics irrespective of free energy or kinetics calculation. $\gamma$ is the bias factor for well-tempered metadynamics.\cite{arpc_meta} 
}
\label{WTmetaD_para}
\end{table}

For high enough supersaturation such as $S_{1}$, $S_{2}$, $S_{3}$ in Table \ref{WTmetaD_para}, we can perform unbiased simulations directly in reasonable computer time, both for the calculation of nucleation kinetics and for feeding stationary density into SGOOP for constructing the RC. However, for lower supersaturations we need to apply enhanced sampling methods since nucleation becomes a rare event. In this work, we use well-tempered metadynamics\cite{wtm, arpc_meta} along a trial RC to obtain preliminary stationary density estimates, and infrequent metadynamics\cite{meta_time,tiwary_rewt} to calculate the kinetics of nucleation.

In metadynamics,\cite{meta_laio,wtm, arpc_meta} the system is encouraged to visit new states by adding a history-dependent Gaussian bias $V(s,t)$ as a function of a biasing variable $s$. Here specifically we use the well-tempered variant of metadynamics in which the height of the Gaussian is tempered through a bias factor each time a point is revisited. The different parameters of the Gaussian bias are listed in Table \ref{WTmetaD_para} in which $h$ is the starting height, $\omega$ is the width, $\Delta t$ is the deposition interval, and $\gamma$ is the bias factor. The final output of a traditional metadynamics run is the free energy along the variable $s$ or along any other degree of freedom which can be expressed as function of the atomic coordinates of the system. In theory, a relation connecting the free energy with the deposited bias can be derived irrespective of the precise choice of biasing variable $s$, which will be asymptotically valid in the limit of long simulation time. In practice however, it helps if $s$ is as close to the true RC as possible.

More recently, a simple extension to well-tempered metadynamics was introduced which allows recovering not just static free energies but also unbiased kinetic information from metadynamics. This protocol has been dubbed ``infrequent metadynamics".\cite{meta_time,tiwary_rewt} The key idea here is that as long as the bias deposition rate is infrequent enough compared to barrier-crossing timescales, in principle we should be able to reweight the biased timescales from well-tempered metadynamics directly to obtain unbiased kinetics through a simple acceleration factor:
\begin{align}
\alpha (t)=\frac{\tau}{\tau_{M}}=\langle e^{\beta V(t)}\rangle
\label{acceleration}
\end{align}
where $\tau$ is the unbiased transition time we seek to learn and $\tau_{M}$ is the biased transition time we actually observe in metadynamics. $V(t)$ is the net bias deposited until time $t$. The central assumption in infrequent metadynamics is that the biasing variable does a good job of timescale separation between time spent in the free energy basin and the time spent during barrier crossing. Thus the need to have a more accurate RC for biasing, that satisfies the criteria of Sec. \ref{sec_rc} becomes even more significant for infrequent metadynamics than in traditional metadynamics. For instance, as we will show in the Results section, infrequent metadynamics becomes more accurate if the biased variable includes all relevant slow modes with long autocorrelation time, and any hidden modes not considered in the biasing variable are as markovian (or quickly decorrelating) as possible. Here we learn such a 1-dimensional RC $\chi$ as a linear combination of our order parameters. Our bias potential then becomes $V(\chi,t)=V(w_{1}n+w_{2}\mu^{2}_{2}+w_{3}\mu^{3}_{3},t)$. The weights of the different order parameters $(w_{1}, w_{2}, w_{3})$, are determined with SGOOP \cite{tiwary2016spectral}.

\section{Results}
\label{results}

\subsection{RC predicted from SGOOP}
\label{rc_definition}
We first describe the RC, as introduced and defined in Sec. \ref{sec_rc}, that we identify for the condensation of a liquid droplet across different supersaturation values. To learn this RC we have used SGOOP, \cite{sgoop,anisod_sgoop} with the stationary probability density $\pi$ estimated through unbiased MD at high supersaturations $S_1$, $S_2$ and preliminary metadynamics\cite{arpc_meta} at low supersaturations $S_4$, $S_4$, $S_5$. The preliminary metadynamics runs were performed biasing $n$. All runs were complemented with short unbiased MD runs (50 ns) for obtaining dynamical constraints for MaxCal. For every supersaturation, SGOOP is initiated from a given choice of trials weights ($w_1,w_2,w_3$) for the RC $\chi$ expressed as $\chi =  w_{1}n+w_{2}\mu^{2}_{2}+w_{3}\mu^{3}_{3}$.

A key question that immediately arises is whether at any given supersaturation $S$ there is a unique RC, or if there are multiple possible combinations of the weights ($w_1,w_2,w_3$) which meet equally well the criteria for an optimal RC described in Sec. \ref{sec_rc}. Yet another question which we ask and answer is how transferable is the RC learnt at one supersaturation $S$ across different values of $S$. To answer the first question, we perform several exhaustive SGOOP trials to estimate the optimized RC, first in the 2-d $(n,\mu_2^2)$ space where we do an explicit grid based search over the full space, and then in the 3-d $(n,\mu_2^2,\mu_3^3)$ space where we start SGOOP from different initial weights. In the latter case, the optimization over weights in SGOOP is performed using a basin hopping algorithm which is a global search algorithm with several stochastic jumps aiding the system from not getting trapped in local minima.

\begin{figure}[h]
  \centering
  \includegraphics[width=8cm]{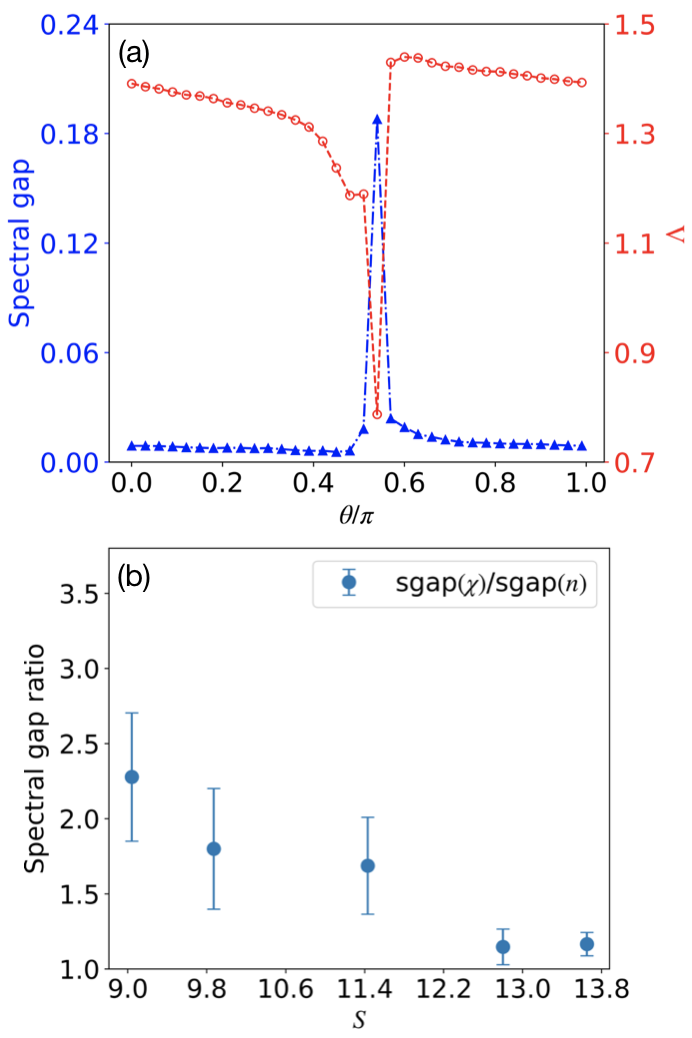} 
  \caption
  { (a) The spectral gap (blue asterisks, left axis) and dynamical prefactor $\Lambda$ (red circles, right axis) of SGOOP transition rate (Eq. \ref{eq_maxcal1}) along different RC $\chi\equiv\cos(\theta)n+\sin(\theta)\mu^{2}_{2}$. Both the maximal spectral gap and minimum $\Lambda$ take place at $\theta=0.5\pi$. (b) Mean spectral gap ratio at five different supersaturation levels $S_1$--$S_5$: For each supersaturation, we averaged the spectral gap ratios calculated from 20 independent biased runs, and the error bars represent the standard error from the averaged results.
}
\label{SGOOP}
\end{figure}

In the 2-d $\chi = w_1 n + w_2 \mu_2^2$ optimization, we do an explicit search among all possible RCs by rotating the putative RC $\chi\equiv\cos(\theta)n+\sin(\theta)\mu^{2}_{2}$ in the  $(n,\mu_2^2)$ space. Here as shown in Fig. \ref{SGOOP}(a) for $S$ = 11.43, we find that the spectral gap profile has a sharp peak when the RC is almost exclusively comprised of $\mu_2^2$, i.e. $\theta \approx \pi/2 $ and $\mu_2^2$ has around 8 times higher weight in the RC than $n$. Such a $\mu_2^2$--heavy RC is obtained irrespective of any $S$ value, showing unequivocally that the second moment $\mu_2^2$ plays a more important role in the RC than $n$ itself. Fig. \ref{SGOOP}(a) also shows the variation of the kinetic pre-factor $\Lambda$ of Eq. \ref{eq_maxcal1} with RC choice, and we will revisit this profile in Sec. \ref{rc_understand}. Next, we perform optimization in the full 3-d $(n,\mu_2^2,\mu_3^3)$ space. Here we find that there are many combinations $(w_{1}, w_{2}, w_{3})$ with similarly enhanced spectral gaps relative to the traditional choice of $\chi=n$, but with a common theme that the second moment and the third moment consistently show up in the optimized RC. 
\begin{figure*}[t]
  \centering
  \includegraphics[width=0.8\textwidth]{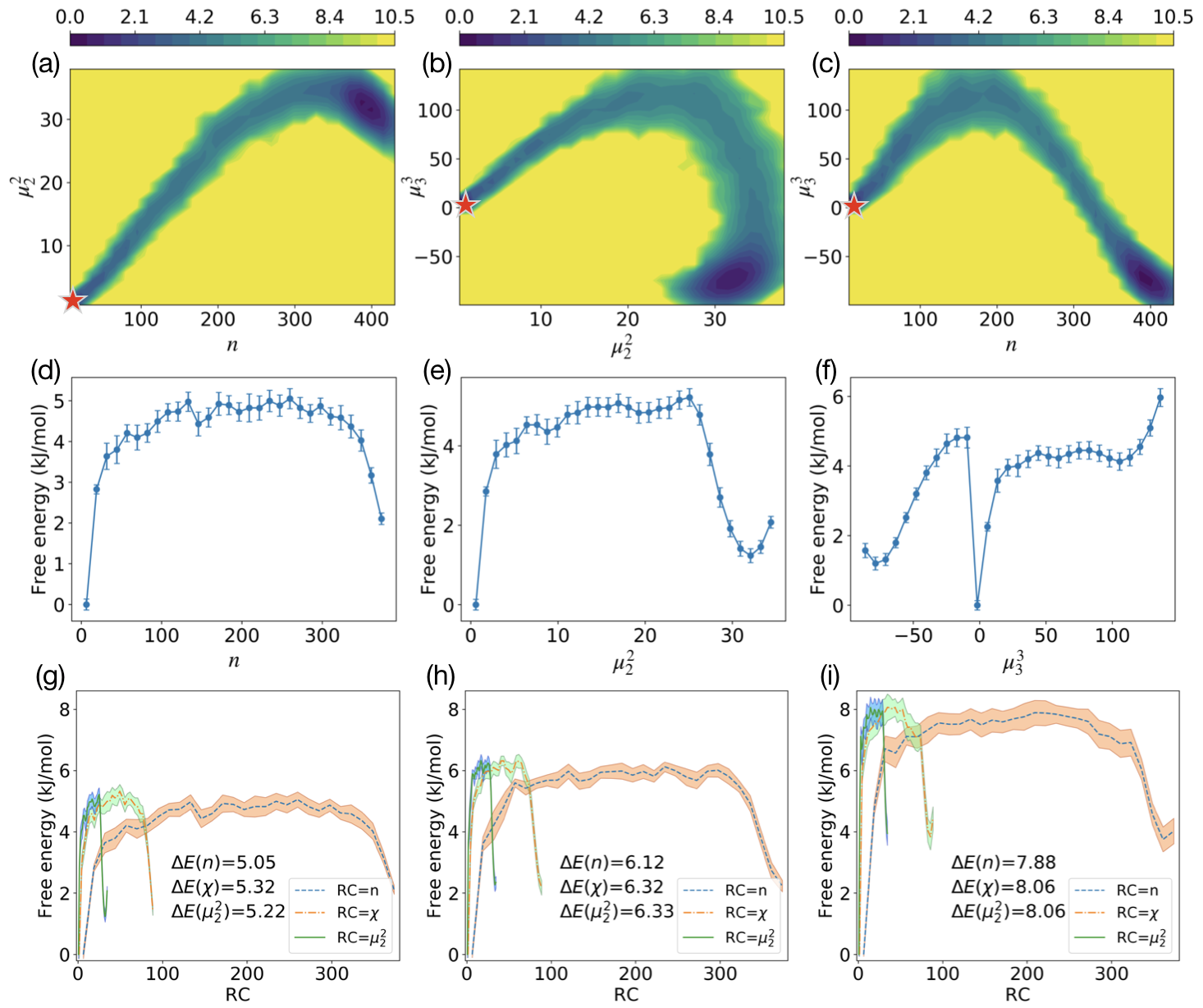} 
  \caption
  {The free energy plots obtained from well-tempered metadynamics biasing along $n$: The top three panels are the 2-d free energy surfaces of (a) (n, $\mu^{2}_{2}$), (b) ($\mu^{2}$, $\mu^{3}_{3}$), and (c) (n,$\mu^{3}$) at supersaturation $S_3$. The starting gaseous state corresponding to each plot is shown with a red star. The middle three panels show the 1-d free energy curves along (d) n, (e) $\mu^{2}_{2}$, and (f) $\mu^{3}_{3}$ respectively. The profiles and the errorbars are calculated from the averages over 10 independent metadynamics runs at supersaturation $S=11.43$. The bottom three panels display the 1-d free energy curves from (g) $S$=11.43, (h) $S$=9.87, and (i) $S$=9.04. In each panel, we show the profile averaged over 10 independent metadynamics runs along n, $\chi$, and $\mu^{2}_{2}$. The regions between errorbars are filled.  The corresponding energy barriers $\Delta E(RC)$ along three different putative RCs are also shown. It can be seen that as $S$ decreases the barrier difference decreases.  All energies are in units of kJ/mol.
}\label{free_energy} 
\end{figure*}

Thus to summarize so far: (a)  RC optimization in $(n,\mu_2^2)$ leads to RC predominantly comprised of $\mu_2^2$, (b) RC optimization in $(n,\mu_2^2,\mu_3^3)$ leads to a RC invariably with weights for all 3 variables, but with multiple local maxima in the spectral gap profile. In other words, the RC is quite degenerate, but considering $\mu_2^2$ and $\mu_3^3$ in the RC is important for a more accurate description of the nucleation process. In the SI we also show results from a full grid search over spectral gaps in the ($w_1,w_2,w_3$) space at at $S$ = 11.43 further illustrating the findings from SGOOP. Here among the first few largest local maxima from 3 different trajectories, we picked ($w_1,w_2,w_3$) = $(0.15, 0.65, -0.15)$ for use in further calculations across all supersaturations $S$. In Fig. \ref{SGOOP} (b), we plot the ratio between the spectral gap along RC = $0.15n+0.65\mu^{2}_{2}-0.15\mu^{3}_{3}$ and that along RC = $n$ at different $S$. As can be seen there, at all $S$ values the optimized RC gives higher spectral gaps than just $n$, and the improvement increases sharply with decreased supersaturation. That is, as the supersaturation decreases the importance of consider shape and density fluctuations in the nuclei become more and more important, which is one of the central findings of this paper. Furthermore, the optimized RC learnt at one supersaturation gives improved spectral gaps at other supersaturations, and hence the RC is transferable across supersaturations. Thus in Sec. \ref{kinetics} we use the RC $\chi = 0.15n+0.65\mu^{2}_{2}-0.15\mu^{3}_{3}$ at all supersaturations for enhanced sampling based calculations of the nucleation rate.

\subsection{Understanding the RC}
\label{rc_understand}
SGOOP optimizes the RC by finding a low-dimensional projection with highest gap between slow and fast processes. In most cases this amounts to selecting a projection with the highest barrier separating the metastable states. To understand if the RC learnt in Sec. \ref{rc_definition} can be attributed to simply barriers in the free energy profile, or if dynamical concerns such as the prefactor $\Lambda$ in Eqs. \ref{eq_maxcal1}--\ref{rate_diffusion} play a role, we construct free energies along various 1-d and 2-d components (totaling 6 combinations) of $(n,\mu_2^2,\mu_3^3)$. These free energies were obtained by running metadynamics with same parameters defined in Table. \ref{WTmetaD_para} and bias potentials added along $n$. We then averaged over 10 independent metadynamics runs with each trajectory reweighted using the free estimator described in Ref. \onlinecite{tiwary_rewt}.

\begin{figure}[t]
  \centering
  \includegraphics[width=8.5cm]{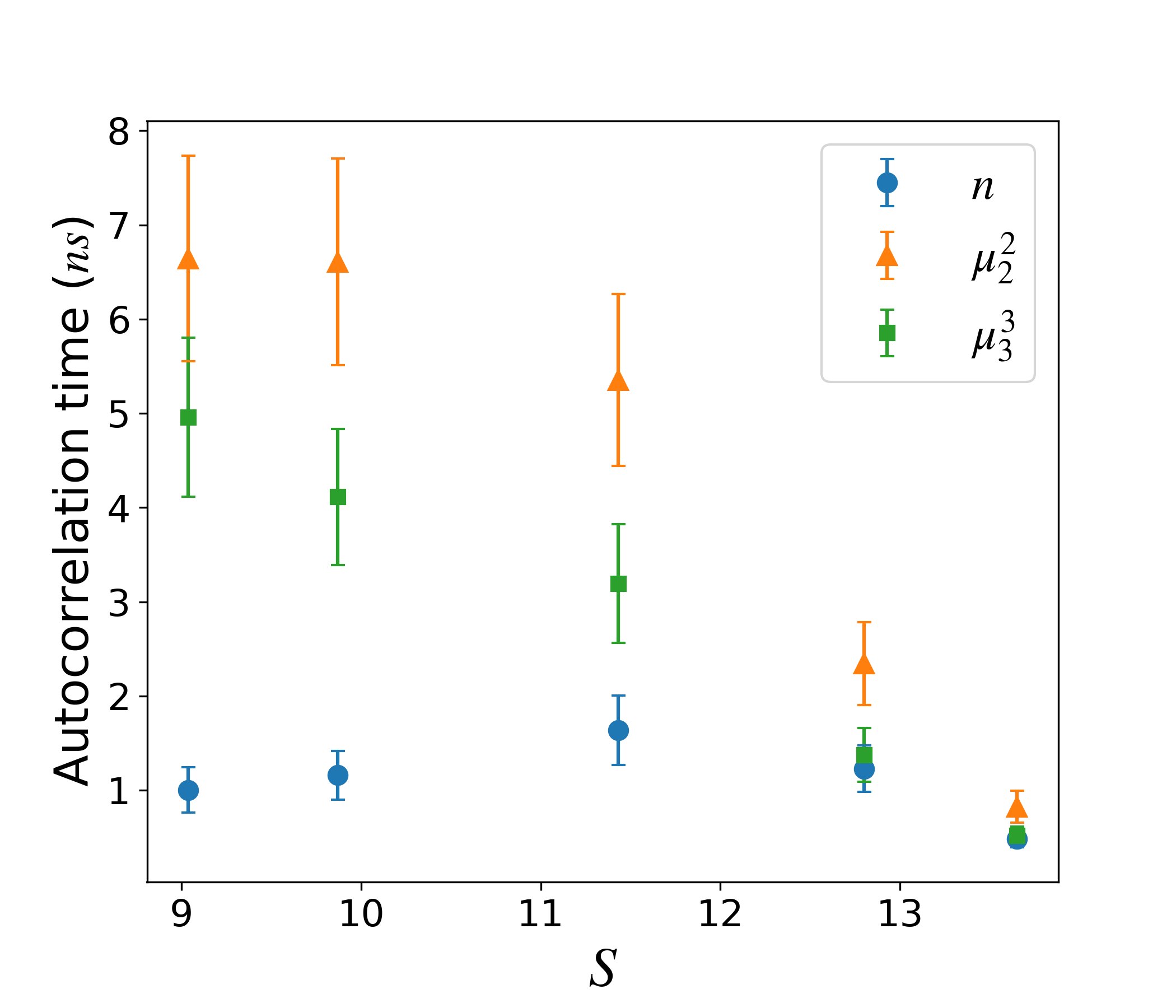}
  \caption
  {The mean autocorrelation times of the order parameters $n$ (blue circles), $\mu_2^2$ (orange triangles), and $\mu^{3}_{3}$ (green squares) calculated from unbiased MD simulations at five different supersaturation levels $S_1$--$S_5$. At each supersaturation level, the calculations from 10 independent runs are averaged. The error bars show the standard error of the averaged results.
}\label{ACT}
\end{figure}

From the various 1-d and 2-d free energy profiles shown in Fig.  \ref{free_energy} (a)-(f) for $S$ = 11.43, it is hard to distinguish between the importance of the various order parameters $n$, $\mu_2^2$,and $\mu_3^3$. The 2-d profiles show that starting from the gas phase (red stars in Fig. \ref{free_energy} (a)-(c)), all three order parameters change in a very correlated manner until the barrier is reached and nucleation is essentially complete ($n$ \textgreater 100). The 1-d free energies along the three order parameters (Fig. \ref{free_energy} (d)-(f)) show that the free energy barrier that needs to be overcome is also very similar for each of the 3 order parameters, though there are some systematic differences which we revisit shortly in Fig. \ref{free_energy} (g)-(i). Comparing the 1-d free energy along $\mu^{3}_{3}$ (Fig. \ref{free_energy} (f)) with the corresponding 2-d free energies (Fig. \ref{free_energy} (b)-(c)), we can see that unlike $n$ and $\mu_2^2$, the 1-d projection along $\mu_3^3$ does a very poor job of describing the pathway in higher dimension space, further justifying our choice of RC $\chi$ in the previous section with higher weight for $\mu_2^2$ than for $\mu_3^3$. In Fig. \ref{free_energy} (g)-(i), we show the free energies along three different RC choices, namely $n$, $\mu_2^2$ and the optimized $\chi=0.15n+0.65\mu^{2}_{2}-0.15\mu^{3}_{3}$, for three different $S$ values. As $S$ is decreased, invariably there is a small but consistent improvement in the barrier height when viewed as function of $\chi$ or $\mu^{2}_{2}$, relative to when viewed as function of $n$. However, firstly this difference is very small (0.25 $kJ$ or 0.1 $k_B T$), and secondly, it appears to get even smaller with decreasing $S$ (Fig. \ref{free_energy} (g)-(i), left to right). Thus, the free energy barrier can not be used to explain the behavior of spectral gap versus supersaturation shown in Fig. \ref{SGOOP} (b). Here we showed that at all supersaturation levels we considered, the spectral gaps of the optimized RC are better than of $n$. It was also pointed out that as supersaturation decreases the spectral gap improvement increases. This tells us that the optimized RC works better at lower supersaturation, which is inconsistent with the change in free energy barriers along different order parameters with supersaturation. 

\begin{figure}[b]
  \centering
  \includegraphics[width=8cm]{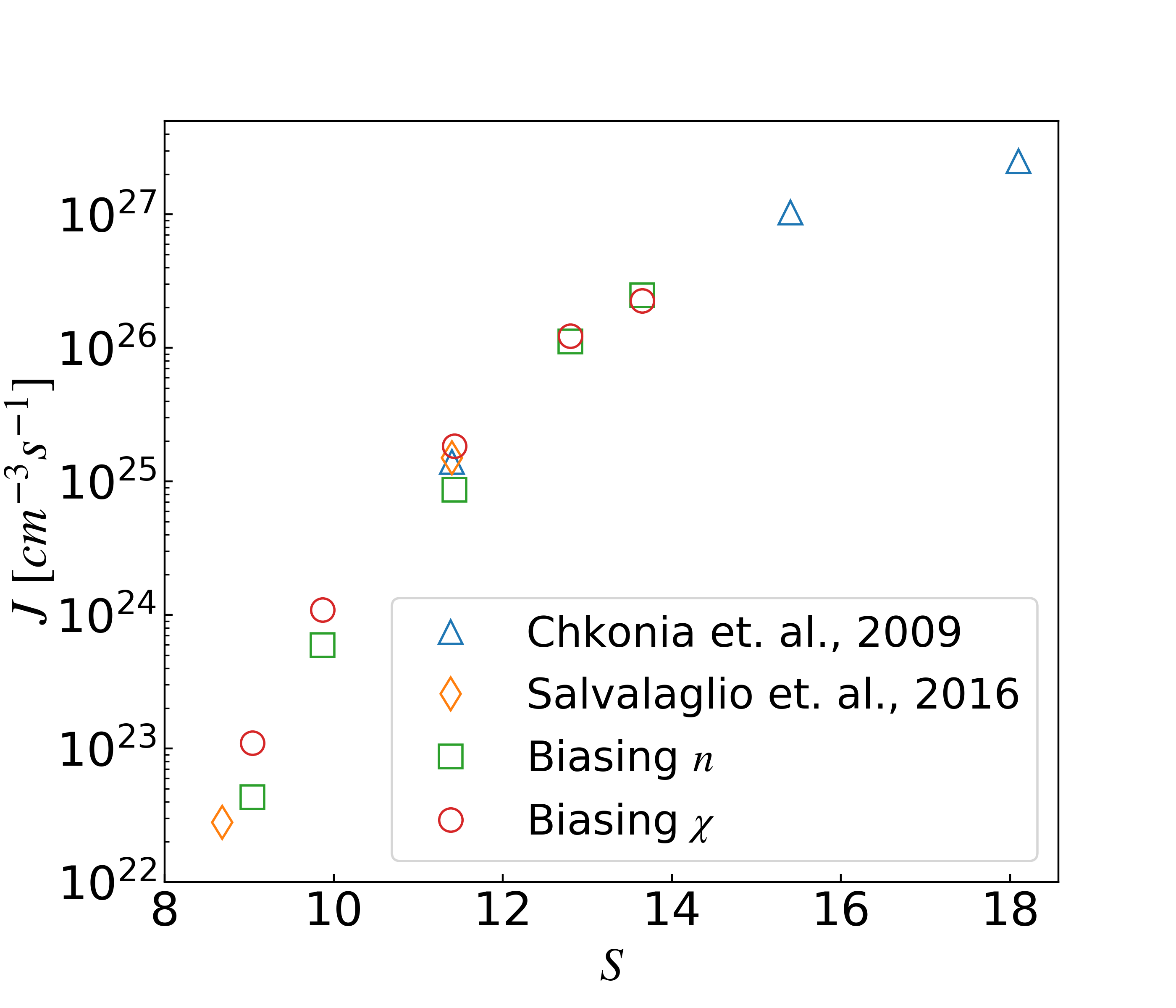} 
  \caption
  { The nucleation rates calculated from the Poisson fits of reweighted nucleation times obtained from infrequent metadynamics biasing along $n$ and $\chi$ (green squares and red circles respectively). The values and their associated error bars are listed in Table \ref{tNucleation}. We also compare our results with previous works from Ref. \onlinecite{chkonia2009evaluating} and Ref. \onlinecite{salvalaglio_argon} (blue triangles and red diamonds respectively).
}\label{J_vs_S} 
\end{figure}

Our next step is therefore explaining why SGOOP finds that $\mu^{2}_{2}$ has a role to play in $\chi$, and why the advantage in considering $\mu^{2}_{2}$ increases with decreasing supersaturation $S$. In Fig. \ref{SGOOP}(a), we provided a profile of how the prefactor $\Lambda$ varied with the RC choice and correspondingly with the spectral gap. It can be seen there that the prefactor $\Lambda$ has a strong inverse correlation with the spectral gap of $\chi$ -- the maximum spectral gap coincides with minimum $\Lambda$. Thus the minuscule increase in barrier height with varying RC is compensated by the slowness of the dynamics along the RC, as captured by $\Lambda$ or the average number of first neighbor transitions in a unit time. 


To gain further insight into this, we calculated time-autocorrelation functions along our three different order parameters (see Fig. \ref{ACT}) as higher autocorrelation time represents slower diffusivity. Our calculations show that $\mu^{2}_{2}$ and $\mu^{3}_{3}$ have longer autocorrelation times than n, and therefore lose memory slower than $n$.\cite{kuipers2009non,ford2004statistical} Furthermore, the increase in autocorrelation times of the two order parameters $\mu^{2}_{2}$ and $\mu^{3}_{3}$ relative to $n$ becomes more and more pronounced as the supersaturation $S$ decreases (Fig. \ref{ACT}). This is in striking contrast to Fig. \ref{free_energy}, where we found an opposite trend looking at the free energy barriers along these order parameters.  

We therefore conclude this section with the observation that anisotropic diffusion in the space of order parameters becomes an important factor in determining the RC especially at lower supersaturations. The longer autocorrelation times are linked to less Markovian behavior, which means $\mu^{2}_{2}$ and $\mu^{3}_{3}$ carry longer memory than $n$.\cite{kuipers2009non,ford2004statistical} Coupled with the finding that all three order parameters have similar barriers in their respective potential of mean force, this means that change in nuclei characteristics such as shape and density become slower as supersaturation $S$ decreases, and it becomes important to explicitly consider this in the construction of a Markovian RC.

\subsection{Nucleation kinetics}
\label{kinetics}
 Now that we have identified an optimized RC $\chi$ = $0.15n+0.65\mu^{2}_{2}-0.15\mu^{3}_{3}$ with improved spectral gap relative to the Frenkel-ten Wolde parameter $n$, we perform two sets of enhanced sampling simulations (specifically, infrequent metadynamics) using $n$ and $\chi$ as biasing variable respectively. We use Eq. \ref{acceleration} to reconstruct the unbiased timescale estimates from these biased runs. At high enough supersaturations we are able to run unbiased MD as well and together with the results of Reguera et al\cite{chkonia2009evaluating} these constitute a valuable set of results to benchmark our findings against. At each supersaturation level, we launched 40 independent metadynamics runs with 20 of them biasing $n$ and the other 20 biasing the optimized RC $\chi$. For each independent run, in order to be able to compare our results with previously published work \cite{chkonia2009evaluating,salvalaglio_argon} we defined the nucleation event as when the coordination number $n$ reaches 30 for the first time. Every independent observation of such an event in terms of its metadynamics time was scaled by the acceleration factor (Eq. \ref{acceleration}) to obtain an unbiased observation of the nucleation time. With these 20 independent estimates of the nucleation time, we can compute the characteristic time (Eq. \ref{poisson_eq}) of observing the first nucleation event $\tau_{N}$ by fitting a Poisson distribution to the statistics, where $\tau_{N}$ is the expected value of the fitted Poisson distribution. The corresponding nucleation rates are then calculated through the formula $J=1/(\tau_{N}V)$ and $J=1/(t_{N}V)$ where V is the volume of system. The results are shown in Table \ref{tNucleation} and in Fig. \ref{J_vs_S}. 

We find that the use of $n$ as a biasing variable in infrequent metadynamics does a remarkably decent job of obtaining nucleation rates (in agreement with the findings of Ref. \onlinecite{salvalaglio_argon}) even with very significant acceleration factors or computational boost relative to unbiased MD. There is nonetheless further improvement that can be obtained in the quality of the nucleation rate if the optimized RC is used instead of $n$, especially as the supersaturation is decreased. As can be seen from Table \ref{tNucleation}, the acceleration factor in metadynamics relative to unbiased MD increases steadily as $S$ decreases, reaching almost four orders of magnitude at the lowest $S$. All reweighted nucleation times, irrespective of whether they came from biasing $n$ or biasing $\chi$ give p-values above the recommended cut-off in the Kolmogorov-Smirnov test from Ref. \onlinecite{pvalue}.  At $S=11.43$, the use of $\chi$ as biasing variable instead of $n$ leads to much better agreement with the unbiased estimate of Reguera et al \cite{chkonia2009evaluating}, as can be seen in  Fig. \ref{J_vs_S}. In general, the characteristic times for nucleation from runs biasing the RC $\chi$ are significantly lower than that those from biasing $n$, and roughly speaking this difference increases as $S$ decreases. In addition to the explicit agreement with unbiased estimate of Reguera et al \cite{chkonia2009evaluating} at $S=11.43$, the lower characteristic time (with similar p-values) can be seen as further evidence of the benefit of biasing $\chi$ instead of $n$. This is because in metadynamics the presence of missing slow degrees of freedom from explicit consideration in the biasing variable typically leads to hysteresis during free energy calculations, or overestimate of the accelerated time through Eq. \ref{acceleration}, as pointed out in Ref. \onlinecite{pvalue} and Ref. \onlinecite{p38}.



\begin{table*}
\renewcommand{\arraystretch}{1.25}
\begin{tabular}{|C{1cm}|C{1cm}|C{5cm}|C{4cm}|C{2cm}|}
\hline
$S$ & RC & $\tau_{N} (s)$ $[p-value]$ & $J(1/\rm{cm}^{3}/s)$ & $\alpha$ \\
\hline
\multirow{2}{*}{$S_1$}
& n & $4.16 \pm 0.45 \times 10^{-9}$ $[0.17]$ & $2.48\pm 0.27\times 10^{26}$ & 1.1 \\ 
& $\chi$ & $4.56 \pm 0.31\times 10^{-9}$ $[0.67]$ & $2.26\pm 0.15\times 10^{26}$ & 1.1 \\ \hline
\multirow{2}{*}{$S_2$}
& n & $8.66 \pm 0.85\times 10^{-9}$ $[0.47]$ & $1.12\pm 0.11\times 10^{26}$ & 2.0 \\ 
& $\chi$ & $7.88 \pm 0.51\times 10^{-9}$ $[0.37]$ & $1.23\pm 0.08\times 10^{26}$ & 1.6 \\ \hline
\multirow{2}{*}{$S_3$}
& n & $1.00 \pm 0.15\times 10^{-7}$ $[0.87]$ & $8.64\pm 1.30\times 10^{24}$ & $4.0\times 10^{1}$ \\ 
& $\chi$ & $0.47 \pm 0.07\times 10^{-7}$ $[0.35]$ & $1.84\pm 0.27\times 10^{25}$ & $6.4\times 10^{1}$ \\ \hline
\multirow{2}{*}{$S_4$}
& n & $1.26 \pm 0.27 \times 10^{-6}$ $[0.64]$ & $5.96\pm 1.28\times 10^{23}$ & $3.7\times 10^{2}$ \\ 
& $\chi$ & $0.69 \pm 0.12 \times 10^{-6}$ $[0.80]$ & $1.09\pm 0.19\times 10^{24}$ & $1.1\times 10^{3}$ \\ \hline
\multirow{2}{*}{$S_5$}
& n & $1.58 \pm 0.19 \times 10^{-5}$ $[0.59]$ & $4.32\pm 0.52\times 10^{22}$ & $5.7\times 10^{3}$ \\ 
& $\chi$ & $0.62 \pm 0.15 \times 10^{-5}$ $[0.13]$ & $1.10\pm 0.27\times 10^{23}$ & $7.8\times 10^{3}$ \\ \hline
\end{tabular}
\caption{The table shows the characteristic nucleation times $\tau_{N}$ by fitting Eq. \ref{poisson_eq} and the corresponding nucleation rates $J$. Results are shown as obtained from the simulations biasing along $n$ as well as biasing along the optimized RC $\chi$. The labels corresponds to the supersaturation levels denoted in Table \ref{WTmetaD_para}. $\alpha$ is the mean acceleration factor for every set of simulations. For the fitted characteristic nucleation times $\tau_{N}$ we have also provided in square brackets the corresponding p-value of the fit when used in Kolmogorov-Smirnov test of Ref. \onlinecite{pvalue}.}
\label{tNucleation}
\end{table*}

\section{Discussion}

In this work, we used new tools \cite{sgoop,meta_time} to revisit a classic problem in nucleation, namely that of the formation of liquid droplet from gaseous precursor as a function of varying driving force for nucleation, namely supersaturation. Our interest was in (a) constructing a Markovian reaction coordinate (RC) for this process, and (b) testing if there is any gain to be had through the use of a more Markovian RC in enhanced sampling calculations of nucleation kinetics. To answer these questions especially at low supersaturations where access to unbiased trajectories of nucleation is difficult (needed by many other RC optimization methods such as Ref. \onlinecite{peters2006obtaining} and Ref. \onlinecite{besthummer_rc}), we use the spectral gap optimization method from Ref. \onlinecite{sgoop} to construct optimized RC from input biased simulations. Our calculations demonstrate unequivocally that it is not sufficient to consider only the typical order parameter used to describe nucleation, namely the number of liquid like atoms in the system. By considering further variables that account for heterogeneity in the system, such as higher moments $\mu^{2}_{2}$ and third moment $\mu^{3}_{3}$ of the distribution of coordination numbers, we could obtain a much more Markovian RC.  Interestingly these various order parameters have nearly identical free energy barriers, and they differ primarily only in associated diffusivities. The importance of these variables further increases with decreasing supersaturation as their associated autocorrelation time increases sharply. In other words, shape and density fluctuations in the nucleating clusters cease to stay rapidly equilibrating variables which can be entirely ignored from a Markovian low-dimensional description of nucleation.  We conclude that diffusion anisotropy plays a more important role at lower $S$, which is supported by our analysis of autocorrelation functions and autocorrelation times. While previous work has demonstrated how infrequent metadynamics can predict nucleation time with only $n$ as the RC, we show in this work that the prediction of nucleation time can be further improved by biasing along an optimized RC. It will be interesting to see if the use of such a more Markovian RC makes improvement in the reliability and efficiency of other enhanced sampling methods such as forward flux sampling.

Finally, it should be mentioned that in this calculation we didn't consider effects due to the finite size of the system, which can be done using the method proposed in Ref. \onlinecite{salvalaglio_argon}, but was not the main objective here. Similarly our findings might change with constant number, pressure and temperature (NPT) simulations. The present work can be redone taking these important nuances into account.
Finally, strictly speaking ours was a model system with model parameters. This work is a proof of principle that ideas such as SGOOP for RC optimization are potentially useful for study of nucleation through enhanced sampling or otherwise. In future we will be extending this work to systems such as crystal nucleation, multiple polymorphs, systems with multiple pathways or multiple species, where there will be even more order parameters to be considered. All of these continue to be very difficult yet important problems for understanding nucleation pathways and rates, and we are hopeful our tools will allow us and others to systematically study these.

Acknowledgment is made to the Donors of the American Chemical Society Petroleum Research Fund for partial support of this research (PRF 60512-DNI6). We also thank Deepthought2, MARCC and XSEDE (projects CHE180007P and CHE180027P) for computational resources used in this work. SGOOP code is available at https://github.com/zwsmith200/SGOOP/

\bibliography{tiwary_references}

\begin{thebibliography}{73}
\expandafter\ifx\csname natexlab\endcsname\relax\def\natexlab#1{#1}\fi
\expandafter\ifx\csname bibnamefont\endcsname\relax
  \def\bibnamefont#1{#1}\fi
\expandafter\ifx\csname bibfnamefont\endcsname\relax
  \def\bibfnamefont#1{#1}\fi
\expandafter\ifx\csname citenamefont\endcsname\relax
  \def\citenamefont#1{#1}\fi
\expandafter\ifx\csname url\endcsname\relax
  \def\url#1{\texttt{#1}}\fi
\expandafter\ifx\csname urlprefix\endcsname\relax\def\urlprefix{URL }\fi
\providecommand{\bibinfo}[2]{#2}
\providecommand{\eprint}[2][]{\url{#2}}

\bibitem[{\citenamefont{Bartels-Rausch}(2013)}]{bartels2013chemistry}
\bibinfo{author}{\bibfnamefont{T.}~\bibnamefont{Bartels-Rausch}},
  \bibinfo{journal}{Nature} \textbf{\bibinfo{volume}{494}}, \bibinfo{pages}{27}
  (\bibinfo{year}{2013}).

\bibitem[{\citenamefont{Murray et~al.}(2010)\citenamefont{Murray, Wilson,
  Dobbie, Cui, Al-Jumur, M{\"o}hler, Schnaiter, Wagner, Benz, Niemand
  et~al.}}]{murray2010heterogeneous}
\bibinfo{author}{\bibfnamefont{B.~J.} \bibnamefont{Murray}},
  \bibinfo{author}{\bibfnamefont{T.~W.} \bibnamefont{Wilson}},
  \bibinfo{author}{\bibfnamefont{S.}~\bibnamefont{Dobbie}},
  \bibinfo{author}{\bibfnamefont{Z.}~\bibnamefont{Cui}},
  \bibinfo{author}{\bibfnamefont{S.~M.} \bibnamefont{Al-Jumur}},
  \bibinfo{author}{\bibfnamefont{O.}~\bibnamefont{M{\"o}hler}},
  \bibinfo{author}{\bibfnamefont{M.}~\bibnamefont{Schnaiter}},
  \bibinfo{author}{\bibfnamefont{R.}~\bibnamefont{Wagner}},
  \bibinfo{author}{\bibfnamefont{S.}~\bibnamefont{Benz}},
  \bibinfo{author}{\bibfnamefont{M.}~\bibnamefont{Niemand}},
  \bibnamefont{et~al.}, \bibinfo{journal}{Nature Geoscience}
  \textbf{\bibinfo{volume}{3}}, \bibinfo{pages}{233} (\bibinfo{year}{2010}).

\bibitem[{\citenamefont{Mazur}(1970)}]{mazur1970cryobiology}
\bibinfo{author}{\bibfnamefont{P.}~\bibnamefont{Mazur}},
  \bibinfo{journal}{Science} \textbf{\bibinfo{volume}{168}},
  \bibinfo{pages}{939} (\bibinfo{year}{1970}).

\bibitem[{\citenamefont{Cox et~al.}(2007)\citenamefont{Cox, Ferris, and
  Thalladi}}]{cox2007selective}
\bibinfo{author}{\bibfnamefont{J.~R.} \bibnamefont{Cox}},
  \bibinfo{author}{\bibfnamefont{L.~A.} \bibnamefont{Ferris}},
  \bibnamefont{and} \bibinfo{author}{\bibfnamefont{V.~R.}
  \bibnamefont{Thalladi}}, \bibinfo{journal}{Angewandte Chemie International
  Edition} \textbf{\bibinfo{volume}{46}}, \bibinfo{pages}{4333}
  (\bibinfo{year}{2007}).

\bibitem[{\citenamefont{Erdemir et~al.}(2007)\citenamefont{Erdemir, Lee, and
  Myerson}}]{erdemir2007polymorph}
\bibinfo{author}{\bibfnamefont{D.}~\bibnamefont{Erdemir}},
  \bibinfo{author}{\bibfnamefont{A.~Y.} \bibnamefont{Lee}}, \bibnamefont{and}
  \bibinfo{author}{\bibfnamefont{A.~S.} \bibnamefont{Myerson}},
  \bibinfo{journal}{Current opinion in drug discovery \& development}
  \textbf{\bibinfo{volume}{10}}, \bibinfo{pages}{746} (\bibinfo{year}{2007}).

\bibitem[{\citenamefont{Sloan}(2003)}]{sloan2003fundamental}
\bibinfo{author}{\bibfnamefont{E.~D.} \bibnamefont{Sloan}},
  \bibinfo{journal}{Nature} \textbf{\bibinfo{volume}{426}},
  \bibinfo{pages}{353} (\bibinfo{year}{2003}).

\bibitem[{\citenamefont{Hammerschmidt}(1934)}]{hammerschmidt1934formation}
\bibinfo{author}{\bibfnamefont{E.}~\bibnamefont{Hammerschmidt}},
  \bibinfo{journal}{Industrial \& Engineering Chemistry}
  \textbf{\bibinfo{volume}{26}}, \bibinfo{pages}{851} (\bibinfo{year}{1934}).

\bibitem[{\citenamefont{Harper et~al.}(1997)\citenamefont{Harper, Lieber, and
  Lansbury~Jr}}]{harper1997atomic}
\bibinfo{author}{\bibfnamefont{J.~D.} \bibnamefont{Harper}},
  \bibinfo{author}{\bibfnamefont{C.~M.} \bibnamefont{Lieber}},
  \bibnamefont{and} \bibinfo{author}{\bibfnamefont{P.~T.}
  \bibnamefont{Lansbury~Jr}}, \bibinfo{journal}{Chemistry \& biology}
  \textbf{\bibinfo{volume}{4}}, \bibinfo{pages}{951} (\bibinfo{year}{1997}).

\bibitem[{\citenamefont{Cohen et~al.}(2013)\citenamefont{Cohen, Linse, Luheshi,
  Hellstrand, White, Rajah, Otzen, Vendruscolo, Dobson, and
  Knowles}}]{cohen2013proliferation}
\bibinfo{author}{\bibfnamefont{S.~I.} \bibnamefont{Cohen}},
  \bibinfo{author}{\bibfnamefont{S.}~\bibnamefont{Linse}},
  \bibinfo{author}{\bibfnamefont{L.~M.} \bibnamefont{Luheshi}},
  \bibinfo{author}{\bibfnamefont{E.}~\bibnamefont{Hellstrand}},
  \bibinfo{author}{\bibfnamefont{D.~A.} \bibnamefont{White}},
  \bibinfo{author}{\bibfnamefont{L.}~\bibnamefont{Rajah}},
  \bibinfo{author}{\bibfnamefont{D.~E.} \bibnamefont{Otzen}},
  \bibinfo{author}{\bibfnamefont{M.}~\bibnamefont{Vendruscolo}},
  \bibinfo{author}{\bibfnamefont{C.~M.} \bibnamefont{Dobson}},
  \bibnamefont{and} \bibinfo{author}{\bibfnamefont{T.~P.}
  \bibnamefont{Knowles}}, \bibinfo{journal}{Proceedings of the National Academy
  of Sciences} \textbf{\bibinfo{volume}{110}}, \bibinfo{pages}{9758}
  (\bibinfo{year}{2013}).

\bibitem[{\citenamefont{Sosso et~al.}(2016)\citenamefont{Sosso, Chen, Cox,
  Fitzner, Pedevilla, Zen, and Michaelides}}]{sosso2016crystal}
\bibinfo{author}{\bibfnamefont{G.~C.} \bibnamefont{Sosso}},
  \bibinfo{author}{\bibfnamefont{J.}~\bibnamefont{Chen}},
  \bibinfo{author}{\bibfnamefont{S.~J.} \bibnamefont{Cox}},
  \bibinfo{author}{\bibfnamefont{M.}~\bibnamefont{Fitzner}},
  \bibinfo{author}{\bibfnamefont{P.}~\bibnamefont{Pedevilla}},
  \bibinfo{author}{\bibfnamefont{A.}~\bibnamefont{Zen}}, \bibnamefont{and}
  \bibinfo{author}{\bibfnamefont{A.}~\bibnamefont{Michaelides}},
  \bibinfo{journal}{Chemical reviews} \textbf{\bibinfo{volume}{116}},
  \bibinfo{pages}{7078} (\bibinfo{year}{2016}).

\bibitem[{\citenamefont{Torrie and Valleau}(1977)}]{torrie1977nonphysical}
\bibinfo{author}{\bibfnamefont{G.~M.} \bibnamefont{Torrie}} \bibnamefont{and}
  \bibinfo{author}{\bibfnamefont{J.~P.} \bibnamefont{Valleau}},
  \bibinfo{journal}{Journal of Computational Physics}
  \textbf{\bibinfo{volume}{23}}, \bibinfo{pages}{187} (\bibinfo{year}{1977}).

\bibitem[{\citenamefont{Kumar et~al.}(1992)\citenamefont{Kumar, Rosenberg,
  Bouzida, Swendsen, and Kollman}}]{kumar1992weighted}
\bibinfo{author}{\bibfnamefont{S.}~\bibnamefont{Kumar}},
  \bibinfo{author}{\bibfnamefont{J.~M.} \bibnamefont{Rosenberg}},
  \bibinfo{author}{\bibfnamefont{D.}~\bibnamefont{Bouzida}},
  \bibinfo{author}{\bibfnamefont{R.~H.} \bibnamefont{Swendsen}},
  \bibnamefont{and} \bibinfo{author}{\bibfnamefont{P.~A.}
  \bibnamefont{Kollman}}, \bibinfo{journal}{Journal of computational chemistry}
  \textbf{\bibinfo{volume}{13}}, \bibinfo{pages}{1011} (\bibinfo{year}{1992}).

\bibitem[{\citenamefont{Laio and Parrinello}(2002)}]{meta_laio}
\bibinfo{author}{\bibfnamefont{A.}~\bibnamefont{Laio}} \bibnamefont{and}
  \bibinfo{author}{\bibfnamefont{M.}~\bibnamefont{Parrinello}},
  \bibinfo{journal}{Proc Natl Acad Sci} \textbf{\bibinfo{volume}{99}},
  \bibinfo{pages}{12562} (\bibinfo{year}{2002}).

\bibitem[{\citenamefont{Barducci et~al.}(2008)\citenamefont{Barducci, Bussi,
  and Parrinello}}]{wtm}
\bibinfo{author}{\bibfnamefont{A.}~\bibnamefont{Barducci}},
  \bibinfo{author}{\bibfnamefont{G.}~\bibnamefont{Bussi}}, \bibnamefont{and}
  \bibinfo{author}{\bibfnamefont{M.}~\bibnamefont{Parrinello}},
  \bibinfo{journal}{Phys Rev Lett} \textbf{\bibinfo{volume}{100}},
  \bibinfo{pages}{020603} (\bibinfo{year}{2008}).

\bibitem[{\citenamefont{van Erp et~al.}(2003)\citenamefont{van Erp, Moroni, and
  Bolhuis}}]{van2003novel}
\bibinfo{author}{\bibfnamefont{T.~S.} \bibnamefont{van Erp}},
  \bibinfo{author}{\bibfnamefont{D.}~\bibnamefont{Moroni}}, \bibnamefont{and}
  \bibinfo{author}{\bibfnamefont{P.~G.} \bibnamefont{Bolhuis}},
  \bibinfo{journal}{The Journal of chemical physics}
  \textbf{\bibinfo{volume}{118}}, \bibinfo{pages}{7762} (\bibinfo{year}{2003}).

\bibitem[{\citenamefont{Moroni et~al.}(2004)\citenamefont{Moroni, van Erp, and
  Bolhuis}}]{moroni2004investigating}
\bibinfo{author}{\bibfnamefont{D.}~\bibnamefont{Moroni}},
  \bibinfo{author}{\bibfnamefont{T.~S.} \bibnamefont{van Erp}},
  \bibnamefont{and} \bibinfo{author}{\bibfnamefont{P.~G.}
  \bibnamefont{Bolhuis}}, \bibinfo{journal}{Physica A: Statistical Mechanics
  and its Applications} \textbf{\bibinfo{volume}{340}}, \bibinfo{pages}{395}
  (\bibinfo{year}{2004}).

\bibitem[{\citenamefont{Allen et~al.}(2006)\citenamefont{Allen, Frenkel, and
  ten Wolde}}]{allen2006simulating}
\bibinfo{author}{\bibfnamefont{R.~J.} \bibnamefont{Allen}},
  \bibinfo{author}{\bibfnamefont{D.}~\bibnamefont{Frenkel}}, \bibnamefont{and}
  \bibinfo{author}{\bibfnamefont{P.~R.} \bibnamefont{ten Wolde}},
  \bibinfo{journal}{The Journal of chemical physics}
  \textbf{\bibinfo{volume}{124}}, \bibinfo{pages}{024102}
  (\bibinfo{year}{2006}).

\bibitem[{\citenamefont{Allen et~al.}(2009)\citenamefont{Allen, Valeriani, and
  ten Wolde}}]{allen2009forward}
\bibinfo{author}{\bibfnamefont{R.~J.} \bibnamefont{Allen}},
  \bibinfo{author}{\bibfnamefont{C.}~\bibnamefont{Valeriani}},
  \bibnamefont{and} \bibinfo{author}{\bibfnamefont{P.~R.} \bibnamefont{ten
  Wolde}}, \bibinfo{journal}{J. Phys.: Cond. Matt.}
  \textbf{\bibinfo{volume}{21}}, \bibinfo{pages}{463102}
  (\bibinfo{year}{2009}).

\bibitem[{\citenamefont{Berezhkovskii and
  Szabo}(2005{\natexlab{a}})}]{berezhkovskii2005one}
\bibinfo{author}{\bibfnamefont{A.}~\bibnamefont{Berezhkovskii}}
  \bibnamefont{and} \bibinfo{author}{\bibfnamefont{A.}~\bibnamefont{Szabo}},
  \bibinfo{journal}{J. Chem. Phys.} \textbf{\bibinfo{volume}{122}},
  \bibinfo{pages}{014503} (\bibinfo{year}{2005}{\natexlab{a}}).

\bibitem[{\citenamefont{Tiwary and Berne}(2016{\natexlab{a}})}]{sgoop}
\bibinfo{author}{\bibfnamefont{P.}~\bibnamefont{Tiwary}} \bibnamefont{and}
  \bibinfo{author}{\bibfnamefont{B.~J.} \bibnamefont{Berne}},
  \bibinfo{journal}{Proc. Natl. Acad. Sci.} \textbf{\bibinfo{volume}{113}},
  \bibinfo{pages}{2839} (\bibinfo{year}{2016}{\natexlab{a}}).

\bibitem[{\citenamefont{Berezhkovskii and
  Szabo}(2005{\natexlab{b}})}]{szabo_anisotropic}
\bibinfo{author}{\bibfnamefont{A.}~\bibnamefont{Berezhkovskii}}
  \bibnamefont{and} \bibinfo{author}{\bibfnamefont{A.}~\bibnamefont{Szabo}},
  \bibinfo{journal}{J. Chem. Phys.} \textbf{\bibinfo{volume}{122}},
  \bibinfo{pages}{014503} (\bibinfo{year}{2005}{\natexlab{b}}).

\bibitem[{\citenamefont{Valsson et~al.}(2016)\citenamefont{Valsson, Tiwary, and
  Parrinello}}]{arpc_meta}
\bibinfo{author}{\bibfnamefont{O.}~\bibnamefont{Valsson}},
  \bibinfo{author}{\bibfnamefont{P.}~\bibnamefont{Tiwary}}, \bibnamefont{and}
  \bibinfo{author}{\bibfnamefont{M.}~\bibnamefont{Parrinello}},
  \bibinfo{journal}{Ann. Rev. Phys. Chem.} \textbf{\bibinfo{volume}{67}},
  \bibinfo{pages}{159} (\bibinfo{year}{2016}).

\bibitem[{\citenamefont{Sarupria and
  Debenedetti}(2012)}]{sarupria2012homogeneous}
\bibinfo{author}{\bibfnamefont{S.}~\bibnamefont{Sarupria}} \bibnamefont{and}
  \bibinfo{author}{\bibfnamefont{P.~G.} \bibnamefont{Debenedetti}},
  \bibinfo{journal}{J. Phys. Chem. Lett.} \textbf{\bibinfo{volume}{3}},
  \bibinfo{pages}{2942} (\bibinfo{year}{2012}).

\bibitem[{\citenamefont{DeFever and Sarupria}(2019)}]{defever2019contour}
\bibinfo{author}{\bibfnamefont{R.~S.} \bibnamefont{DeFever}} \bibnamefont{and}
  \bibinfo{author}{\bibfnamefont{S.}~\bibnamefont{Sarupria}},
  \bibinfo{journal}{J. Chem. Phys.} \textbf{\bibinfo{volume}{150}},
  \bibinfo{pages}{024103} (\bibinfo{year}{2019}).

\bibitem[{\citenamefont{Dellago et~al.}(1998)\citenamefont{Dellago, Bolhuis,
  and Chandler}}]{dellago1998efficient}
\bibinfo{author}{\bibfnamefont{C.}~\bibnamefont{Dellago}},
  \bibinfo{author}{\bibfnamefont{P.~G.} \bibnamefont{Bolhuis}},
  \bibnamefont{and} \bibinfo{author}{\bibfnamefont{D.}~\bibnamefont{Chandler}},
  \bibinfo{journal}{The Journal of chemical physics}
  \textbf{\bibinfo{volume}{108}}, \bibinfo{pages}{9236} (\bibinfo{year}{1998}).

\bibitem[{\citenamefont{Bolhuis et~al.}(1998)}]{bolhuis1998sampling}
\bibinfo{author}{\bibfnamefont{P.}~\bibnamefont{Bolhuis}} \bibnamefont{et~al.},
  \bibinfo{journal}{Faraday Discussions} \textbf{\bibinfo{volume}{110}},
  \bibinfo{pages}{421} (\bibinfo{year}{1998}).

\bibitem[{\citenamefont{Chkonia et~al.}(2009)\citenamefont{Chkonia, W{\"o}lk,
  Strey, Wedekind, and Reguera}}]{chkonia2009evaluating}
\bibinfo{author}{\bibfnamefont{G.}~\bibnamefont{Chkonia}},
  \bibinfo{author}{\bibfnamefont{J.}~\bibnamefont{W{\"o}lk}},
  \bibinfo{author}{\bibfnamefont{R.}~\bibnamefont{Strey}},
  \bibinfo{author}{\bibfnamefont{J.}~\bibnamefont{Wedekind}}, \bibnamefont{and}
  \bibinfo{author}{\bibfnamefont{D.}~\bibnamefont{Reguera}},
  \bibinfo{journal}{The Journal of chemical physics}
  \textbf{\bibinfo{volume}{130}}, \bibinfo{pages}{064505}
  (\bibinfo{year}{2009}).

\bibitem[{\citenamefont{Salvalaglio et~al.}(2016)\citenamefont{Salvalaglio,
  Tiwary, Maggioni, Mazzotti, and Parrinello}}]{salvalaglio_argon}
\bibinfo{author}{\bibfnamefont{M.}~\bibnamefont{Salvalaglio}},
  \bibinfo{author}{\bibfnamefont{P.}~\bibnamefont{Tiwary}},
  \bibinfo{author}{\bibfnamefont{G.~M.} \bibnamefont{Maggioni}},
  \bibinfo{author}{\bibfnamefont{M.}~\bibnamefont{Mazzotti}}, \bibnamefont{and}
  \bibinfo{author}{\bibfnamefont{M.}~\bibnamefont{Parrinello}},
  \bibinfo{journal}{J. Chem. Phys.} \textbf{\bibinfo{volume}{145}},
  \bibinfo{pages}{211925} (\bibinfo{year}{2016}).

\bibitem[{\citenamefont{Kuipers and Barkema}(2009)}]{kuipers2009non}
\bibinfo{author}{\bibfnamefont{J.}~\bibnamefont{Kuipers}} \bibnamefont{and}
  \bibinfo{author}{\bibfnamefont{G.}~\bibnamefont{Barkema}},
  \bibinfo{journal}{Physical Review E} \textbf{\bibinfo{volume}{79}},
  \bibinfo{pages}{062101} (\bibinfo{year}{2009}).

\bibitem[{\citenamefont{Ford}(2004)}]{ford2004statistical}
\bibinfo{author}{\bibfnamefont{I.}~\bibnamefont{Ford}},
  \bibinfo{journal}{Proceedings of the Institution of Mechanical Engineers,
  Part C: Journal of Mechanical Engineering Science}
  \textbf{\bibinfo{volume}{218}}, \bibinfo{pages}{883} (\bibinfo{year}{2004}).

\bibitem[{\citenamefont{Tiwary and Berne}(2017{\natexlab{a}})}]{anisod_sgoop}
\bibinfo{author}{\bibfnamefont{P.}~\bibnamefont{Tiwary}} \bibnamefont{and}
  \bibinfo{author}{\bibfnamefont{B.~J.} \bibnamefont{Berne}},
  \bibinfo{journal}{J. Chem. Phys.} \textbf{\bibinfo{volume}{147}},
  \bibinfo{pages}{152701} (\bibinfo{year}{2017}{\natexlab{a}}).

\bibitem[{\citenamefont{Tiwary and
  Berne}(2016{\natexlab{b}})}]{sgoop_fullerene}
\bibinfo{author}{\bibfnamefont{P.}~\bibnamefont{Tiwary}} \bibnamefont{and}
  \bibinfo{author}{\bibfnamefont{B.~J.} \bibnamefont{Berne}},
  \bibinfo{journal}{J. Chem. Phys.} \textbf{\bibinfo{volume}{145}},
  \bibinfo{eid}{054113} (\bibinfo{year}{2016}{\natexlab{b}}),
  \urlprefix\url{http://scitation.aip.org/content/aip/journal/jcp/145/5/10.1063/1.4959969}.

\bibitem[{\citenamefont{Smith et~al.}(2018)\citenamefont{Smith, Pramanik, Tsai,
  and Tiwary}}]{multisgoop}
\bibinfo{author}{\bibfnamefont{Z.}~\bibnamefont{Smith}},
  \bibinfo{author}{\bibfnamefont{D.}~\bibnamefont{Pramanik}},
  \bibinfo{author}{\bibfnamefont{S.-T.} \bibnamefont{Tsai}}, \bibnamefont{and}
  \bibinfo{author}{\bibfnamefont{P.}~\bibnamefont{Tiwary}},
  \bibinfo{journal}{J. Chem. Phys.} \textbf{\bibinfo{volume}{149}},
  \bibinfo{pages}{234105} (\bibinfo{year}{2018}).

\bibitem[{\citenamefont{Weber and Volmer}(1926)}]{weber1926keimbildung}
\bibinfo{author}{\bibfnamefont{A.}~\bibnamefont{Weber}} \bibnamefont{and}
  \bibinfo{author}{\bibfnamefont{M.}~\bibnamefont{Volmer}},
  \bibinfo{journal}{Zeitschrift f{\"u}r Physikalische Chemie}
  \textbf{\bibinfo{volume}{119}}, \bibinfo{pages}{277} (\bibinfo{year}{1926}).

\bibitem[{\citenamefont{Farkas}(1927)}]{farkas1927keimbildungsgeschwindigkeit}
\bibinfo{author}{\bibfnamefont{L.}~\bibnamefont{Farkas}},
  \bibinfo{journal}{Zeitschrift f{\"u}r physikalische Chemie}
  \textbf{\bibinfo{volume}{125}}, \bibinfo{pages}{236} (\bibinfo{year}{1927}).

\bibitem[{\citenamefont{Becker and D{\"o}ring}(1935)}]{becker1935kinetische}
\bibinfo{author}{\bibfnamefont{R.}~\bibnamefont{Becker}} \bibnamefont{and}
  \bibinfo{author}{\bibfnamefont{W.}~\bibnamefont{D{\"o}ring}},
  \bibinfo{journal}{Annalen der Physik} \textbf{\bibinfo{volume}{416}},
  \bibinfo{pages}{719} (\bibinfo{year}{1935}).

\bibitem[{\citenamefont{Zeldovich}(1943)}]{zeldovich1943theory}
\bibinfo{author}{\bibfnamefont{Y.~B.} \bibnamefont{Zeldovich}},
  \bibinfo{journal}{Acta Physicochem., USSR} \textbf{\bibinfo{volume}{18}},
  \bibinfo{pages}{1} (\bibinfo{year}{1943}).

\bibitem[{\citenamefont{Maris}(2006)}]{maris2006introduction}
\bibinfo{author}{\bibfnamefont{H.~J.} \bibnamefont{Maris}},
  \bibinfo{journal}{Comptes Rendus Physique} \textbf{\bibinfo{volume}{7}},
  \bibinfo{pages}{946} (\bibinfo{year}{2006}).

\bibitem[{\citenamefont{Kalikmanov}(2013)}]{Kalikmanov2013}
\bibinfo{author}{\bibfnamefont{V.~I.} \bibnamefont{Kalikmanov}},
  \emph{\bibinfo{title}{Classical Nucleation Theory}}
  (\bibinfo{publisher}{Springer Netherlands}, \bibinfo{address}{Dordrecht},
  \bibinfo{year}{2013}), pp. \bibinfo{pages}{17--41}, ISBN
  \bibinfo{isbn}{978-90-481-3643-8},
  \urlprefix\url{https://doi.org/10.1007/978-90-481-3643-8_3}.

\bibitem[{\citenamefont{Berezhkovskii and
  Szabo}(2011{\natexlab{a}})}]{szabo_timescale}
\bibinfo{author}{\bibfnamefont{A.}~\bibnamefont{Berezhkovskii}}
  \bibnamefont{and} \bibinfo{author}{\bibfnamefont{A.}~\bibnamefont{Szabo}},
  \bibinfo{journal}{J. Chem. Phys.} \textbf{\bibinfo{volume}{135}},
  \bibinfo{pages}{074108} (\bibinfo{year}{2011}{\natexlab{a}}).

\bibitem[{\citenamefont{van~der Zwan and Hynes}(1982)}]{van1982reactive}
\bibinfo{author}{\bibfnamefont{G.}~\bibnamefont{van~der Zwan}}
  \bibnamefont{and} \bibinfo{author}{\bibfnamefont{J.~T.} \bibnamefont{Hynes}},
  \bibinfo{journal}{J. Chem. Phys.} \textbf{\bibinfo{volume}{77}},
  \bibinfo{pages}{1295} (\bibinfo{year}{1982}).

\bibitem[{\citenamefont{Hynes}(1985)}]{hynes1985chemical}
\bibinfo{author}{\bibfnamefont{J.~T.} \bibnamefont{Hynes}},
  \bibinfo{journal}{Annual Review of Physical Chemistry}
  \textbf{\bibinfo{volume}{36}}, \bibinfo{pages}{573} (\bibinfo{year}{1985}).

\bibitem[{\citenamefont{Peters et~al.}(2013)\citenamefont{Peters, Bolhuis,
  Mullen, and Shea}}]{peters2013reaction}
\bibinfo{author}{\bibfnamefont{B.}~\bibnamefont{Peters}},
  \bibinfo{author}{\bibfnamefont{P.~G.} \bibnamefont{Bolhuis}},
  \bibinfo{author}{\bibfnamefont{R.~G.} \bibnamefont{Mullen}},
  \bibnamefont{and} \bibinfo{author}{\bibfnamefont{J.-E.} \bibnamefont{Shea}},
  \bibinfo{journal}{J. Chem. Phys.} \textbf{\bibinfo{volume}{138}},
  \bibinfo{pages}{054106} (\bibinfo{year}{2013}).

\bibitem[{\citenamefont{Tiwary and Parrinello}(2013)}]{meta_time}
\bibinfo{author}{\bibfnamefont{P.}~\bibnamefont{Tiwary}} \bibnamefont{and}
  \bibinfo{author}{\bibfnamefont{M.}~\bibnamefont{Parrinello}},
  \bibinfo{journal}{Phys. Rev. Lett.} \textbf{\bibinfo{volume}{111}},
  \bibinfo{pages}{230602} (\bibinfo{year}{2013}).

\bibitem[{\citenamefont{Rao et~al.}(1978)\citenamefont{Rao, Berne, and
  Kalos}}]{rao1978computer}
\bibinfo{author}{\bibfnamefont{M.}~\bibnamefont{Rao}},
  \bibinfo{author}{\bibfnamefont{B.}~\bibnamefont{Berne}}, \bibnamefont{and}
  \bibinfo{author}{\bibfnamefont{M.}~\bibnamefont{Kalos}},
  \bibinfo{journal}{The Journal of Chemical Physics}
  \textbf{\bibinfo{volume}{68}}, \bibinfo{pages}{1325} (\bibinfo{year}{1978}).

\bibitem[{\citenamefont{Wang et~al.}(2007)\citenamefont{Wang, Gould, and
  Klein}}]{wang2007homogeneous}
\bibinfo{author}{\bibfnamefont{H.}~\bibnamefont{Wang}},
  \bibinfo{author}{\bibfnamefont{H.}~\bibnamefont{Gould}}, \bibnamefont{and}
  \bibinfo{author}{\bibfnamefont{W.}~\bibnamefont{Klein}},
  \bibinfo{journal}{Physical Review E} \textbf{\bibinfo{volume}{76}},
  \bibinfo{pages}{031604} (\bibinfo{year}{2007}).

\bibitem[{\citenamefont{Kalikmanov et~al.}(2008)\citenamefont{Kalikmanov,
  W{\"o}lk, and Kraska}}]{kalikmanov2008argon}
\bibinfo{author}{\bibfnamefont{V.}~\bibnamefont{Kalikmanov}},
  \bibinfo{author}{\bibfnamefont{J.}~\bibnamefont{W{\"o}lk}}, \bibnamefont{and}
  \bibinfo{author}{\bibfnamefont{T.}~\bibnamefont{Kraska}},
  \bibinfo{journal}{The Journal of chemical physics}
  \textbf{\bibinfo{volume}{128}}, \bibinfo{pages}{124506}
  (\bibinfo{year}{2008}).

\bibitem[{\citenamefont{ten Wolde and Frenkel}(1997)}]{ten1997enhancement}
\bibinfo{author}{\bibfnamefont{P.~R.} \bibnamefont{ten Wolde}}
  \bibnamefont{and} \bibinfo{author}{\bibfnamefont{D.}~\bibnamefont{Frenkel}},
  \bibinfo{journal}{Science} \textbf{\bibinfo{volume}{277}},
  \bibinfo{pages}{1975} (\bibinfo{year}{1997}).

\bibitem[{\citenamefont{Trudu et~al.}(2006)\citenamefont{Trudu, Donadio, and
  Parrinello}}]{trudu2006freezing}
\bibinfo{author}{\bibfnamefont{F.}~\bibnamefont{Trudu}},
  \bibinfo{author}{\bibfnamefont{D.}~\bibnamefont{Donadio}}, \bibnamefont{and}
  \bibinfo{author}{\bibfnamefont{M.}~\bibnamefont{Parrinello}},
  \bibinfo{journal}{Physical review letters} \textbf{\bibinfo{volume}{97}},
  \bibinfo{pages}{105701} (\bibinfo{year}{2006}).

\bibitem[{\citenamefont{Moroni et~al.}(2005)\citenamefont{Moroni, Ten~Wolde,
  and Bolhuis}}]{moroni2005interplay}
\bibinfo{author}{\bibfnamefont{D.}~\bibnamefont{Moroni}},
  \bibinfo{author}{\bibfnamefont{P.~R.} \bibnamefont{Ten~Wolde}},
  \bibnamefont{and} \bibinfo{author}{\bibfnamefont{P.~G.}
  \bibnamefont{Bolhuis}}, \bibinfo{journal}{Physical review letters}
  \textbf{\bibinfo{volume}{94}}, \bibinfo{pages}{235703}
  (\bibinfo{year}{2005}).

\bibitem[{\citenamefont{Kwon et~al.}(2015)\citenamefont{Kwon, Krylova,
  Phillips, Klie, Chattopadhyay, Shibata, Bunel, Liu, Prakapenka, Lee
  et~al.}}]{kwon2015heterogeneous}
\bibinfo{author}{\bibfnamefont{S.~G.} \bibnamefont{Kwon}},
  \bibinfo{author}{\bibfnamefont{G.}~\bibnamefont{Krylova}},
  \bibinfo{author}{\bibfnamefont{P.~J.} \bibnamefont{Phillips}},
  \bibinfo{author}{\bibfnamefont{R.~F.} \bibnamefont{Klie}},
  \bibinfo{author}{\bibfnamefont{S.}~\bibnamefont{Chattopadhyay}},
  \bibinfo{author}{\bibfnamefont{T.}~\bibnamefont{Shibata}},
  \bibinfo{author}{\bibfnamefont{E.~E.} \bibnamefont{Bunel}},
  \bibinfo{author}{\bibfnamefont{Y.}~\bibnamefont{Liu}},
  \bibinfo{author}{\bibfnamefont{V.~B.} \bibnamefont{Prakapenka}},
  \bibinfo{author}{\bibfnamefont{B.}~\bibnamefont{Lee}}, \bibnamefont{et~al.},
  \bibinfo{journal}{Nature materials} \textbf{\bibinfo{volume}{14}},
  \bibinfo{pages}{215} (\bibinfo{year}{2015}).

\bibitem[{\citenamefont{Zhou et~al.}(2019)\citenamefont{Zhou, Yang, Yang, Kim,
  Yuan, Tian, Ophus, Sun, Schmid, Nathanson et~al.}}]{zhou2019observing}
\bibinfo{author}{\bibfnamefont{J.}~\bibnamefont{Zhou}},
  \bibinfo{author}{\bibfnamefont{Y.}~\bibnamefont{Yang}},
  \bibinfo{author}{\bibfnamefont{Y.}~\bibnamefont{Yang}},
  \bibinfo{author}{\bibfnamefont{D.~S.} \bibnamefont{Kim}},
  \bibinfo{author}{\bibfnamefont{A.}~\bibnamefont{Yuan}},
  \bibinfo{author}{\bibfnamefont{X.}~\bibnamefont{Tian}},
  \bibinfo{author}{\bibfnamefont{C.}~\bibnamefont{Ophus}},
  \bibinfo{author}{\bibfnamefont{F.}~\bibnamefont{Sun}},
  \bibinfo{author}{\bibfnamefont{A.~K.} \bibnamefont{Schmid}},
  \bibinfo{author}{\bibfnamefont{M.}~\bibnamefont{Nathanson}},
  \bibnamefont{et~al.}, \bibinfo{journal}{Nature}
  \textbf{\bibinfo{volume}{570}}, \bibinfo{pages}{500} (\bibinfo{year}{2019}).

\bibitem[{\citenamefont{ten Wolde and Frenkel}(1998)}]{ten1998computer}
\bibinfo{author}{\bibfnamefont{P.~R.} \bibnamefont{ten Wolde}}
  \bibnamefont{and} \bibinfo{author}{\bibfnamefont{D.}~\bibnamefont{Frenkel}},
  \bibinfo{journal}{The Journal of chemical physics}
  \textbf{\bibinfo{volume}{109}}, \bibinfo{pages}{9901} (\bibinfo{year}{1998}).

\bibitem[{\citenamefont{Tribello et~al.}(2017)\citenamefont{Tribello, Giberti,
  Sosso, Salvalaglio, and Parrinello}}]{tribello2017analyzing}
\bibinfo{author}{\bibfnamefont{G.~A.} \bibnamefont{Tribello}},
  \bibinfo{author}{\bibfnamefont{F.}~\bibnamefont{Giberti}},
  \bibinfo{author}{\bibfnamefont{G.~C.} \bibnamefont{Sosso}},
  \bibinfo{author}{\bibfnamefont{M.}~\bibnamefont{Salvalaglio}},
  \bibnamefont{and}
  \bibinfo{author}{\bibfnamefont{M.}~\bibnamefont{Parrinello}},
  \bibinfo{journal}{Journal of chemical theory and computation}
  \textbf{\bibinfo{volume}{13}}, \bibinfo{pages}{1317} (\bibinfo{year}{2017}).

\bibitem[{\citenamefont{Tribello et~al.}(2010)\citenamefont{Tribello, Ceriotti,
  and Parrinello}}]{tribello2010self}
\bibinfo{author}{\bibfnamefont{G.~A.} \bibnamefont{Tribello}},
  \bibinfo{author}{\bibfnamefont{M.}~\bibnamefont{Ceriotti}}, \bibnamefont{and}
  \bibinfo{author}{\bibfnamefont{M.}~\bibnamefont{Parrinello}},
  \bibinfo{journal}{Proc. Natl. Acad. Sci.} \textbf{\bibinfo{volume}{107}},
  \bibinfo{pages}{17509} (\bibinfo{year}{2010}).

\bibitem[{\citenamefont{Kathmann et~al.}(2004)\citenamefont{Kathmann, Schenter,
  and Garrett}}]{kathmann2004multicomponent}
\bibinfo{author}{\bibfnamefont{S.~M.} \bibnamefont{Kathmann}},
  \bibinfo{author}{\bibfnamefont{G.~K.} \bibnamefont{Schenter}},
  \bibnamefont{and} \bibinfo{author}{\bibfnamefont{B.~C.}
  \bibnamefont{Garrett}}, \bibinfo{journal}{J. Chem. Phys.}
  \textbf{\bibinfo{volume}{120}}, \bibinfo{pages}{9133} (\bibinfo{year}{2004}).

\bibitem[{\citenamefont{Resnick}(2013)}]{resnick2013adventures}
\bibinfo{author}{\bibfnamefont{S.~I.} \bibnamefont{Resnick}},
  \emph{\bibinfo{title}{Adventures in stochastic processes}}
  (\bibinfo{publisher}{Springer Science \& Business Media},
  \bibinfo{year}{2013}).

\bibitem[{\citenamefont{Salvalaglio et~al.}(2014)\citenamefont{Salvalaglio,
  Tiwary, and Parrinello}}]{pvalue}
\bibinfo{author}{\bibfnamefont{M.}~\bibnamefont{Salvalaglio}},
  \bibinfo{author}{\bibfnamefont{P.}~\bibnamefont{Tiwary}}, \bibnamefont{and}
  \bibinfo{author}{\bibfnamefont{M.}~\bibnamefont{Parrinello}},
  \bibinfo{journal}{J. Chem. Theor. Comp.} \textbf{\bibinfo{volume}{10}},
  \bibinfo{pages}{1420} (\bibinfo{year}{2014}).

\bibitem[{\citenamefont{Bussi et~al.}(2007)\citenamefont{Bussi, Donadio, and
  Parrinello}}]{bussi2007canonical}
\bibinfo{author}{\bibfnamefont{G.}~\bibnamefont{Bussi}},
  \bibinfo{author}{\bibfnamefont{D.}~\bibnamefont{Donadio}}, \bibnamefont{and}
  \bibinfo{author}{\bibfnamefont{M.}~\bibnamefont{Parrinello}},
  \bibinfo{journal}{The Journal of chemical physics}
  \textbf{\bibinfo{volume}{126}}, \bibinfo{pages}{014101}
  (\bibinfo{year}{2007}).

\bibitem[{\citenamefont{Lindahl et~al.}(2001)\citenamefont{Lindahl, Hess, and
  Van Der~Spoel}}]{lindahl2001gromacs}
\bibinfo{author}{\bibfnamefont{E.}~\bibnamefont{Lindahl}},
  \bibinfo{author}{\bibfnamefont{B.}~\bibnamefont{Hess}}, \bibnamefont{and}
  \bibinfo{author}{\bibfnamefont{D.}~\bibnamefont{Van Der~Spoel}},
  \bibinfo{journal}{Molecular modeling annual} \textbf{\bibinfo{volume}{7}},
  \bibinfo{pages}{306} (\bibinfo{year}{2001}).

\bibitem[{\citenamefont{Massimiliano~Bonomi}(2019)}]{plumed2019nature}
\bibinfo{author}{\bibfnamefont{C.~C. e.~a.} \bibnamefont{Massimiliano~Bonomi},
  \bibfnamefont{Giovanni~Bussi}}, \bibinfo{journal}{Nature methods}
  \textbf{\bibinfo{volume}{16}}, \bibinfo{pages}{670} (\bibinfo{year}{2019}).

\bibitem[{\citenamefont{Berezhkovskii and
  Szabo}(2011{\natexlab{b}})}]{berezhkovskii2011time}
\bibinfo{author}{\bibfnamefont{A.}~\bibnamefont{Berezhkovskii}}
  \bibnamefont{and} \bibinfo{author}{\bibfnamefont{A.}~\bibnamefont{Szabo}},
  \bibinfo{journal}{The Journal of chemical physics}
  \textbf{\bibinfo{volume}{135}}, \bibinfo{pages}{074108}
  (\bibinfo{year}{2011}{\natexlab{b}}).

\bibitem[{\citenamefont{Dixit et~al.}(2015)\citenamefont{Dixit, Jain, Stock,
  and Dill}}]{dixit2015inferring}
\bibinfo{author}{\bibfnamefont{P.~D.} \bibnamefont{Dixit}},
  \bibinfo{author}{\bibfnamefont{A.}~\bibnamefont{Jain}},
  \bibinfo{author}{\bibfnamefont{G.}~\bibnamefont{Stock}}, \bibnamefont{and}
  \bibinfo{author}{\bibfnamefont{K.~A.} \bibnamefont{Dill}},
  \bibinfo{journal}{J. Chem. Theor. Comp.} \textbf{\bibinfo{volume}{11}},
  \bibinfo{pages}{5464} (\bibinfo{year}{2015}).

\bibitem[{\citenamefont{Jaynes}(1980)}]{jaynes_caliber}
\bibinfo{author}{\bibfnamefont{E.~T.} \bibnamefont{Jaynes}},
  \bibinfo{journal}{Ann. Rev. Phys. Chem.} \textbf{\bibinfo{volume}{31}},
  \bibinfo{pages}{579} (\bibinfo{year}{1980}).

\bibitem[{\citenamefont{Press\'e et~al.}(2013)\citenamefont{Press\'e, Ghosh,
  Lee, and Dill}}]{caliber1}
\bibinfo{author}{\bibfnamefont{S.}~\bibnamefont{Press\'e}},
  \bibinfo{author}{\bibfnamefont{K.}~\bibnamefont{Ghosh}},
  \bibinfo{author}{\bibfnamefont{J.}~\bibnamefont{Lee}}, \bibnamefont{and}
  \bibinfo{author}{\bibfnamefont{K.~A.} \bibnamefont{Dill}},
  \bibinfo{journal}{Rev. Mod. Phys.} \textbf{\bibinfo{volume}{85}},
  \bibinfo{pages}{1115} (\bibinfo{year}{2013}),
  \urlprefix\url{http://link.aps.org/doi/10.1103/RevModPhys.85.1115}.

\bibitem[{\citenamefont{Dixit et~al.}(2018)\citenamefont{Dixit, Wagoner,
  Weistuch, Press{\'e}, Ghosh, and Dill}}]{dixit2018perspective}
\bibinfo{author}{\bibfnamefont{P.~D.} \bibnamefont{Dixit}},
  \bibinfo{author}{\bibfnamefont{J.}~\bibnamefont{Wagoner}},
  \bibinfo{author}{\bibfnamefont{C.}~\bibnamefont{Weistuch}},
  \bibinfo{author}{\bibfnamefont{S.}~\bibnamefont{Press{\'e}}},
  \bibinfo{author}{\bibfnamefont{K.}~\bibnamefont{Ghosh}}, \bibnamefont{and}
  \bibinfo{author}{\bibfnamefont{K.~A.} \bibnamefont{Dill}},
  \bibinfo{journal}{The Journal of chemical physics}
  \textbf{\bibinfo{volume}{148}}, \bibinfo{pages}{010901}
  (\bibinfo{year}{2018}).

\bibitem[{\citenamefont{Tiwary and Parrinello}(2014)}]{tiwary_rewt}
\bibinfo{author}{\bibfnamefont{P.}~\bibnamefont{Tiwary}} \bibnamefont{and}
  \bibinfo{author}{\bibfnamefont{M.}~\bibnamefont{Parrinello}},
  \bibinfo{journal}{J. Phys. Chem. B} \textbf{\bibinfo{volume}{119}},
  \bibinfo{pages}{736} (\bibinfo{year}{2014}).

\bibitem[{\citenamefont{Tiwary and
  Berne}(2017{\natexlab{b}})}]{tiwary2017predicting}
\bibinfo{author}{\bibfnamefont{P.}~\bibnamefont{Tiwary}} \bibnamefont{and}
  \bibinfo{author}{\bibfnamefont{B.}~\bibnamefont{Berne}},
  \bibinfo{journal}{The Journal of chemical physics}
  \textbf{\bibinfo{volume}{147}}, \bibinfo{pages}{152701}
  (\bibinfo{year}{2017}{\natexlab{b}}).

\bibitem[{\citenamefont{Bicout and Szabo}(1998)}]{szabo_bicout}
\bibinfo{author}{\bibfnamefont{D.}~\bibnamefont{Bicout}} \bibnamefont{and}
  \bibinfo{author}{\bibfnamefont{A.}~\bibnamefont{Szabo}}, \bibinfo{journal}{J.
  Chem. Phys.} \textbf{\bibinfo{volume}{109}}, \bibinfo{pages}{2325}
  (\bibinfo{year}{1998}).

\bibitem[{\citenamefont{Tiwary and
  Berne}(2016{\natexlab{c}})}]{tiwary2016spectral}
\bibinfo{author}{\bibfnamefont{P.}~\bibnamefont{Tiwary}} \bibnamefont{and}
  \bibinfo{author}{\bibfnamefont{B.}~\bibnamefont{Berne}},
  \bibinfo{journal}{Proceedings of the National Academy of Sciences}
  \textbf{\bibinfo{volume}{113}}, \bibinfo{pages}{2839}
  (\bibinfo{year}{2016}{\natexlab{c}}).

\bibitem[{\citenamefont{Casasnovas et~al.}(2017)\citenamefont{Casasnovas,
  Limongelli, Tiwary, Carloni, and Parrinello}}]{p38}
\bibinfo{author}{\bibfnamefont{R.}~\bibnamefont{Casasnovas}},
  \bibinfo{author}{\bibfnamefont{V.}~\bibnamefont{Limongelli}},
  \bibinfo{author}{\bibfnamefont{P.}~\bibnamefont{Tiwary}},
  \bibinfo{author}{\bibfnamefont{P.}~\bibnamefont{Carloni}}, \bibnamefont{and}
  \bibinfo{author}{\bibfnamefont{M.}~\bibnamefont{Parrinello}},
  \bibinfo{journal}{J. Am. Chem. Soc.} \textbf{\bibinfo{volume}{139}},
  \bibinfo{pages}{4780} (\bibinfo{year}{2017}).

\bibitem[{\citenamefont{Peters and Trout}(2006)}]{peters2006obtaining}
\bibinfo{author}{\bibfnamefont{B.}~\bibnamefont{Peters}} \bibnamefont{and}
  \bibinfo{author}{\bibfnamefont{B.~L.} \bibnamefont{Trout}},
  \bibinfo{journal}{The Journal of chemical physics}
  \textbf{\bibinfo{volume}{125}}, \bibinfo{pages}{054108}
  (\bibinfo{year}{2006}).

\bibitem[{\citenamefont{Best and Hummer}(2005)}]{besthummer_rc}
\bibinfo{author}{\bibfnamefont{R.~B.} \bibnamefont{Best}} \bibnamefont{and}
  \bibinfo{author}{\bibfnamefont{G.}~\bibnamefont{Hummer}},
  \bibinfo{journal}{Proc. Natl. Acad. Sci.} \textbf{\bibinfo{volume}{102}},
  \bibinfo{pages}{6732} (\bibinfo{year}{2005}).

\end{thebibliography}

	\end{document}